\documentclass[aps,pra,twocolumn,superscriptaddress]{revtex4-1} 
\usepackage{graphicx}
\usepackage{color}
\usepackage{amssymb}
\usepackage{amsmath}
\usepackage[squaren]{SIunits}
\usepackage{comment}

\usepackage{capt-of} 

\usepackage[squaren]{SIunits}

\definecolor{myurlcolor}{rgb}{0,0,0.7}
\definecolor{myrefcolor}{rgb}{0.8,0,0}
\usepackage[unicode=true,pdfusetitle, bookmarks=false,bookmarksnumbered=false,
bookmarksopen=false, breaklinks=false,pdfborder={0 0 0},backref=false,
colorlinks=true, linkcolor=myrefcolor,citecolor=myurlcolor,urlcolor=myurlcolor]{hyperref}

\newcommand{\Intens}{{\cal I}}
\newcommand{\INucl}{I}

\newcommand{\nulight}{\nu_\t{light}}
\newcommand{\PRLsection}[1]{\emph{#1}.---}

\newcommand{\gP}{g_{\textrm{P}}} 
\newcommand{\gD}{g_{\textrm{D}}} 

\newcommand{\vecb}[2]{\mathbf{#1}{_{#2}}}  
\newcommand{\estRJM}[3]{\tilde{\mathbf{#1}}{_{#2 | #3}}} 
\newcommand{\covRJM}[2]{\Sigma_{#1 | #2}} 
\newcommand{\mean}[1]{\mathrm{E}\!\left[#1\right]}

\newcommand{\nd}{\cN}

\renewcommand{\t}[1]{\mathrm{#1}}

\newcommand{\vB}[1]{\mathbf{#1}} 
\newcommand{\mB}[1]{\mathbf{#1}} 
\newcommand{\est}[1]{\tilde{#1}} 
\newcommand{\innov}{\tilde{\vB{y}}} 
\newcommand{\cov}{\mB{\Sigma}} 
\newcommand{\covSSp}{\mB{\check{\Sigma}}_{\t{ss}}} 
\newcommand{\covSSpRF}{\mB{\check{\Sigma}}_{\t{ss}}^{\t{RF}}} 
\newcommand{\covSSu}{\mB{\Sigma}_{\t{ss}}} 
\newcommand{\covSSuRF}{\mB{\Sigma}_{\t{ss}}^{\t{RF}}} 
\newcommand{\mzero}{\mB{0}} 
\newcommand{\e}{\mathrm{e}}
\renewcommand{\d}{\mathrm{d}}

\newcommand{\ot}{\otimes}

\newcommand{\tr}[1]{\textrm{Tr}\!\left\{#1\right\}}
\newcommand{\cE}{\mathcal{E}}
\newcommand{\cN}{\mathcal{N}}

\newcommand{\eref}[1]{\eqref{#1}}
\newcommand{\eqnref}[1]{Eq.~\eqref{#1}}
\newcommand{\eqnsref}[2]{Eqs.~\eqref{#1} and \eqref{#2}}
\newcommand{\figref}[1]{Fig.~\ref{#1}}
\newcommand{\tabref}[1]{Table~\ref{#1}}

\newcommand{\appref}[1]{App.~\ref{#1}}

\newcommand{\refcite}[1]{Ref.~\cite{#1}}

\newcommand{\BE}{\begin{equation}}
\newcommand{\EE}{\end{equation}}
\newcommand{\AL}[1]{\begin{align}#1\end{align}}

\newcommand{\BEA}{\begin{eqnarray}}
\newcommand{\EEA}{\end{eqnarray}}

\definecolor{mygreen}{rgb}{0,0.0,0}
\definecolor{myred}{rgb}{0.75,0,0}
\definecolor{myblue}{rgb}{0,0,0.0}
\definecolor{mymagenta}{cmyk}{0,1,0,0.12}
\definecolor{mycyan}{cmyk}{1,0,0,0.12}
\definecolor{myorange}{rgb}{1,0.5,0}
\newcommand{\btext}[1]{{\color{myblue}#1}}

\newcommand{\gtext}[1]{{\color{mygreen}#1}}

\usepackage[normalem]{ulem}

\newcommand{\ICFO}{\affiliation{ICFO--Institut de Ciencies Fotoniques, The Barcelona Institute of Science and Technology, 08860 Castelldefels (Barcelona), Spain}}
\newcommand{\ICREA}{\affiliation{ICREA--Instituci\'o Catalana de Recerca i Estudis Avan\c{c}ats, 08010 Barcelona, Spain}}

\begin{document}

\newcommand{\MyTitle}{Signal tracking beyond the time resolution of an atomic sensor by Kalman filtering}
\title{\MyTitle} 
\author{Ricardo Jim\'{e}nez-Mart\'{i}nez}
\ICFO
\author{Jan Ko\l{}ody\'{n}ski}
\ICFO
\author{Charikleia Troullinou}
\ICFO
\author{Vito Giovanni Lucivero}
\ICFO
\author{Jia Kong}
\ICFO
\author{Morgan W. Mitchell}
\ICFO
\ICREA

\date{July 27, 2017}

\begin{abstract}
We study causal waveform estimation (tracking) of time-varying signals in a paradigmatic atomic sensor, an alkali vapor monitored by Faraday rotation probing. We use Kalman filtering, which optimally tracks known linear Gaussian stochastic processes, to estimate stochastic input signals that we generate by optical pumping. Comparing the known input to the estimates, we confirm the accuracy of the atomic statistical model and the reliability of the Kalman filter, allowing recovery of waveform details far briefer than the sensor's intrinsic time resolution.  With proper filter choice, we obtain similar benefits when tracking partially-known and non-Gaussian signal processes, as are found in most practical sensing applications. The method evades the trade-off between sensitivity and time resolution in coherent sensing.
\end{abstract}
\maketitle

\PRLsection{Introduction}%
Extremely precise sensors, e.g., atomic clocks \cite{ludlow2015}, magnetometers \cite{KominisN2003}, and gravitational-wave detectors \cite{LIGOPRL2016a} employ a two-stage transducing architecture. A quantity of interest, e.g., electromagnetic field or gravitational-wave strain, coherently drives a well-isolated \emph{sensing component}, e.g., the suspended mirrors of an interferometer or the spins of an atomic ensemble. The sensing component is non-destructively measured or ``read out'' by a second, \emph{meter component}, often an optical beam. The two-stage architecture isolates the sensor component, enabling high coherence and high sensitivity \cite{KominisN2003}, but also complicates the signal interpretation. In atomic sensors, for example, the slow spin-response, as well as intrinsic noises in spin orientation and in the readout, can distort and mask the signal \cite{ShahPRL2010}.

One compelling application of such sensors is estimation of time-varying signals, e.g., gravitational \cite{LIGOPRL2016a} or biomagnetic events \cite{SanderBOE2012}. For this application, the central statistical problem is \emph{waveform estimation} \cite{Tsang2009,TsangPRL2011}. In control applications \cite{GeremiaS2004,Yonezawa2012,Wieczorek2015}, the estimation must also be performed in real time \cite{Hochberg2012,Lien2016}, as when a spectroscopy signal is fed back to a local oscillator in an atomic clock \cite{ludlow2015}.

Tools from Bayesian statistics \cite{vanTrees2013,vanTrees2007bayesian} provide \btext{a} natural framework for waveform estimation 
with multi-stage sensors. Of particular interest is the \emph{Kalman filter} (KF)  that \btext{provides fast and causal estimation} \cite{Kalman1960,Kalman1961}. \btext{For linear Gaussian models, KF estimates are moreover optimal} (i.e., with minimum mean squared error) and provide \btext{a full statistical} description of the waveform. Sophisticated methods extend the KF technique to more general problems \cite{vanTrees2007bayesian}. Even \btext{when not optimal,} the KF is often applied for its simplicity, versatility and controllability \cite{slam2006,Cassola2012}.

To date KFs have been experimentally implemented in optical sensors:~to estimate the phase of a light beam \cite{Yonezawa2012}, to track an external force applied to a mirror in a quantum-enhanced interferometer \cite{Iwasawa2013}, and to estimate in real time the quantum state of an optomechanical oscillator \cite{Wieczorek2015}. Application to atomic sensors promises to benefit applications in magnetometry \cite{KominisN2003,Lucivero2014}, gyroscopy \cite{KornackPRL2005}, gravimetry \cite{AngelisMST2009}, optical NMR \cite{Jimenez2014a}, fundamental physics \cite{SmiciklasPRL2011}, and quantum communications \cite{julsgaard2004}. Here, we \btext{demonstrate} KFs in an archetypal two-stage atomic sensor:~an atomic spin ensemble read-out via the optical Faraday effect. 

Using spin polarization by optical pumping we apply known waveforms, which enables us to compare the KF estimates against the true value of the signal. In this way we first verify the accuracy of the statistical model underlying the KF and a major expected benefit of the KF approach---optimal waveform estimation including signal components faster than the intrinsic temporal resolution of the sensor. We also study estimation of waveforms with dynamics only partially known to the observer. The optimality studied in prior works \cite{Yonezawa2012,Iwasawa2013,Wieczorek2015} is not present in this scenario \cite{vanTrees2013}, which includes many important sensing problems \cite{SanderBOE2012,alem2015fetal,bar2004}. For appropriately-constructed KFs, we nonetheless observe advantages in speed and sensitivity, making the KF attractive for general-purpose atomic sensing. 
\begin{figure*}[!t]
	\centering
	\includegraphics[width=\textwidth]{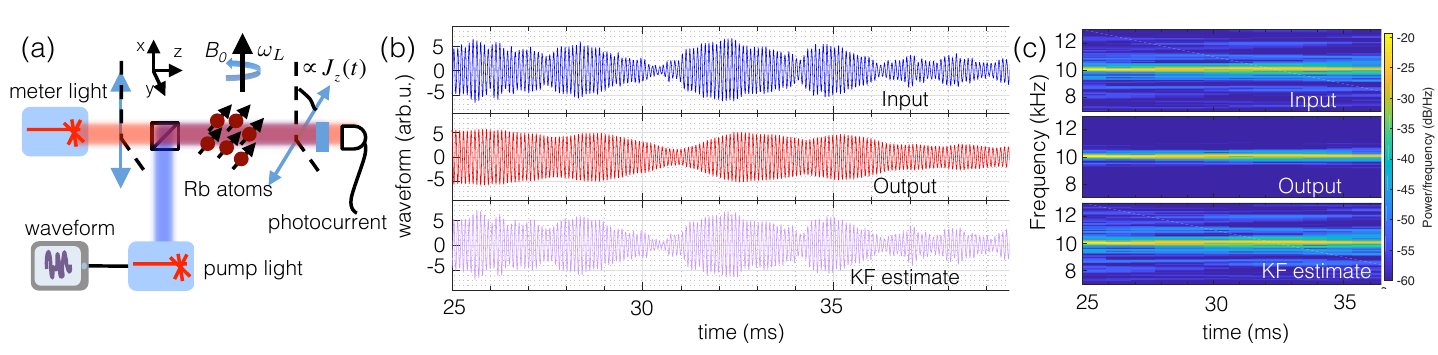}
\caption{%
(a):~An ensemble of $^\t{87}$Rb atoms precesses at the Larmor frequency $\omega_L  = 2 \pi \times \unit{10}{\kilo\hertz}$ defined by an external magnetic field $B_0$. The spin $z$-component, $J_z(t)$, is driven by a circularly polarized light-beam (pump) carrying a waveform to be estimated. A second laser-beam (meter) is used for a polarimetry measurement producing a photocurrent that is  proportional to $J_z(t)$  plus shot noise. Transmitted pump light is blocked by a dichroic filter (shown in blue). (b):~A representative applied waveform (input) along with the corresponding measured photocurrent (output) and the recovered waveform (KF estimate). (c):~Spectrograms of input, output and KF estimate showing that rapidly-varying features of the input are suppressed in the output yet are recovered in the KF estimate.}
\label{fig:setup}
\end{figure*}

\PRLsection{Two-stage sensor}%
The sensor is depicted in \figref{fig:setup}(a). Its \emph{sensing component} consists of an atomic ensemble exhibiting a total spin $J$, whose components 
$J_{i}(t)\!=\!\t{Tr}\{\rho(t)\hat{J}_{i}\}$ with $i \in \{x,y,z\}$ are determined by the collective spin-operators, $\hat{J}_{i}$, and the ensemble state at time $t$, $\rho(t)$. The dynamics of $J$ includes: precession about $x$ at the Larmor (angular) frequency $\omega_L$ due to a known magnetic field $B_0$, coupling to the drive \textit{signal} $\cE(t)$ applied using circularly polarized pump-light along $z$, as well as relaxation and noise processes associated with atomic collisions, optical depolarization and transit-time broadening (all effectively characterised by the $T_2$-parameter and stochastic fluctuations measured via noise spectroscopy \cite{Lucivero2016a,Lucivero2016b}). The $J_z(t)$ spin of the ensemble is read out by the \emph{meter component} of the sensor---a linearly-polarized off-resonance light beam propagating along $z$---which experiences Faraday rotation by an angle proportional to $J_z(t)$ yielding the detected photocurrent $I(t)$ subject to shot noise. 

As shown in \figref{fig:setup}(b), rapidly-varying features in an applied waveform appear distorted in the output, due to the slow response of the atoms. See, e.g., the dip at \unit{30.5}{\milli\second}, which appears only after a delay of \mbox{$\approx\!\unit{0.5}{\milli\second}$} with considerable loss of fast features. Despite this, a KF (described below) tracks these features as they occur in real time. To achieve these results the KF relies on a statistical model for spin, waveform, and detection dynamics.

\PRLsection{Statistical model}%
We describe the dynamics of the spin components $\vB{j}_t=[J_{y}(t),J_{z}(t)]^T$ using the linear Gaussian model of Refs.~\cite{Lucivero2016a,Lucivero2016b}, which after translating from frequency to time domain reads:
\BE
\d \vB{j}_t = \begin{bmatrix}-\frac{1}{T_2} & \omega_\t{L} \\ -\omega_\t{L} & -\frac{1}{T_2}\end{bmatrix}\vB{j}_t \,\d t
+ \begin{bmatrix} 0\\  \cE(t) \end{bmatrix} \d t + \d \vB{w}_t^{(\t J)},
\label{eq:dyns_atoms}
\EE
where the spin-noise vector $\d\vB{w}_t^{(\t J)}=[\d w_y(t),\d w_z(t)]^T$ 
describes independent stochastic increments $\d w_\alpha(t) = \sqrt{Q_{\alpha}} \d W_{\alpha}(t)$ ($\alpha \in \{y,z\}$) obeying Gaussian 
white-noise statistics that we denote using the normal distribution
$\d w_\alpha(t) \sim \cN(0,Q_\alpha\d t)$ with mean $\mean{ \d w_\alpha(t) } = 0$ and variance $\mean{\d w_\alpha(t)\d w_\beta(t)} = \delta_{\alpha\beta}Q_\alpha \d t$, \gtext{where the scalar strength $Q_\alpha>0$ is determined experimentally (see \cite{suppmat}).}

The signal in \eref{eq:dyns_atoms}, $\cE(t)=[\gP\cos(\omega_{\t{P}}t),\gP\sin(\omega_{\t{P}}t)]\cdot\vB{q}_t$, contains the quadrature components $\vB{q}_t=[q(t),p(t)]^T$ that we aim to estimate, while the carrier-frequency $\omega_{\t{P}}$ and coupling constant $\gP$ are known parameters \cite{suppmat}. In the validation experiment of this work, we drive the atoms with a signal whose quadratures are described by independent 
Ornstein-Uhlenbeck (OU) processes \cite{gardiner1985} with correlation time $\kappa^{-1}$:
\BE
\d \vB{q}_t = -\kappa\,\vB{q}_t \d t + \d \vB{w}_t^{(\t q)}, 
\label{eq:dyns_pump} 
\EE
where $\d \vB{w}_t^{(\t q)}=[\d w_q(t),\d w_p(t)]^T$ denotes the noise vector of the quadrature components containing independent stochastic increments that are defined in an analogous manner to the spin-noise vector in \eqnref{eq:dyns_atoms}. 

We monitor the atomic spins via optical Faraday rotation of the meter beam \cite{suppmat}. As discussed in previous works \cite{Lucivero2016a,Lucivero2016b}, the photocurrent $I(t)$ produced by this detection process is subjected to Gaussian white-noise 
of the light (i.e., optical shot noise). In our experiments we sample the photocurrent at finite-time intervals, i.e., at $t_k\!=\!k\,\Delta$ 
with integer $k$ and sampling period $\Delta$. To account for this fact, we describe the sensor output by a discrete-time 
stochastic equation of the form \cite{suppmat}:
\BE
I(t_k) = \gD\, J_z(t_k)+\xi_\t{D}(t_k),
\label{eq:I_meas}
\EE
where $\gD$ denotes the transduction constant in our experimental setup and 
$\xi_\t{D}(t_k)\sim\cN(0,R^\Delta)$ represents the white-noise of each observation, 
with variance $R^\Delta = R/\Delta$ dictated by the power-spectral-density, $R$, 
of the optical shot-noise and the sampling period, $\Delta$. 

\PRLsection{Kalman Filter}%
In state estimation problems, the goal is to construct an estimator $\est{\vB{x}}_t$, that optimally tracks the state $\vB{x}_t$ of a system which, despite possessing known dynamics, cannot be directly measured
due to detection noise and its intrinsic fluctuations  \cite{vanTrees2013}.
 \btext{For linear-Gaussian systems}, the optimal estimator---minimising the mean squared error---is provided by the Kalman Filter (KF) \cite{Kalman1960,Kalman1961}. For time-continuous processes integration-based versions of the KF are favored, e.g., 
Kalman-Bucy filters \cite{Tsang2009PRA}. However, as the output of our sensor is sampled at discrete times, we focus 
on \btext{its} continuous-discrete version \cite{bar2004,suppmat} applicable to dynamics described by 
the general state-space model of linear systems \cite{bar2004}:
\AL{
    	\d\vB{x}_t  &=\mB{F}_t \vB{x}_t \d t+\d\vB{w}_t, \label{eq:dyns_state}\\
	\vB{z}_{k} &=\mB{H}_{k}\vB{x}_k + \vB{v}_k, \label{eq:dyns_obs}
}
where $\vB{x}_t$ and $\vB{z}_k\equiv\vB{z}_{t_k}$ are the state and observation vectors describ\btext{ing} the system and measurement processes, respectively. For the atomic sensor under study, we define the state vector $\vB{x}_t = \vB{j}_t\oplus \vB{q}_t$, so that the \emph{system dynamics} encompasses the evolution of transversal spin-components and 
signal quadratures, i.e., \eqnsref{eq:dyns_atoms}{eq:dyns_pump}, respectively. The stochastic increment in \eqnref{eq:dyns_state} is then formed by a direct sum, $\d\vB{w}_t=\d \vB{w}_t^{(\t J)} \oplus \d \vB{w}_t^{(\t q)}$, of the corresponding spin- and quadrature-noise vectors and satisfies $\mean{\d \vB{w}_t}=\vB{0}$,  $\mean{\d \vB{w}_t \d \vB{w}_t^T} = \mB{Q}\,\d t$ with $\mB{Q}=\t{diag}\{Q_y,Q_z,Q_q,Q_p\}$ being its $4\!\times\!4$ diagonal covariance matrix. An explicit expression for the matrix $\mB{F}_t$ applicable to our atomic sensor can be found in \refcite{suppmat}. The photocurrent \eref{eq:I_meas}, on the other hand, constitutes the (scalar) \emph{measurement model} \eref{eq:dyns_obs} with $\vB{z}_k \!\equiv\! z_k\!=\!I(t_k)$, $\mB{H}_k \equiv \mB{H}= [0,\gD,0,0]$, and  $\vB{v}_k \!\equiv\! v_k\!=\!\xi_\t{D}(t_k)$.

In the continuous-discrete KF the \emph{estimate}, $\est{\vB{x}}_t$, and its \emph{error covariance matrix}, $\cov_t=\mean{(\vB{x}_t -\est{\vB{x}}_t)(\vB{x}_t-\est{\vB{x}}_t )^{T}}$, are constructed in a two-step procedure \cite{Jazwinski1970,vanTrees2013}. First, their values at $t_k$, $\est{\vB{x}}_{k|k-1}$ and $\cov_{k|k-1}$, are \emph{predicted} conditioned on the previous instance, $\est{\vB{x}}_{k-1|k-1}$ and $\cov_{k-1|k-1}$, as follows:
\AL{	
	 \est{\vB{x}}_{{k|k-1}}  &= \mB{\Phi}_{k,k-1}\est{\vB{x}}_{k-1|k-1},  \\
	 \cov_{k|k-1} &=  \vecb{\Phi}{k,k-1}\cov_{k-1|k-1}\mB{\Phi}_{k,k-1}^T + \mB{Q}_k^\Delta, \label{eq:cov_pred}
} 
where $\mB{\Phi}_{k,k-1}$ is the \emph{transition matrix} describing the solution of the dynamical model \eqref{eq:dyns_state} \cite{Kalman1960}. $\mB{Q}_k^\Delta$ is then
the effective covariance matrix of the system noise, $\mB{Q}$, that 
now adequately accounts for the finite sampling period, $\Delta$, of the measurement \cite{suppmat}.
Second, the \emph{update} step is performed according to the rule:
\AL{	\est{\vB{x}}_{k|k}  &= \est{\vB{x}}_{k|k-1} + \mB{K}_{k}\,\innov_k, \\
	\cov_{k|k} &=  \left(\openone -\mB{K}_{k}\mB{H}_{k} \right)\cov_{k|k-1}, \label{Eq:kf-cov-main}
}
after computing the \emph{innovation} $\innov_k$ and the \emph{Kalman gain} $\mB{K}_k$ that depend on the ``fresh''{} observation $\vB{z}_k$, i.e.,
\BE	
\innov_k =\vB{z}_k-\tilde{\vB{z}}_k,
\qquad
\mB{K}_{k} =\cov_{k|k-1}\mB{H}_{k}^{T} \vB{S}_{k}^{-1},
\EE
where $\tilde{\vB{z}}_k=\vecb{H}{k}\est{\vB{x}}_{k|k-1}$ represents the KF estimate of the $k$th observation, whose precision is then quantified by the covariance matrix:
\BE
	\vecb{S}{k} = \mean{\innov_k \innov_k^T} = \mB{R}^{\Delta} + \mB{H}_{k}\covRJM{k}{k-1}\vB{H}_{k}^T.
	\label{Eq:kf-out-cov}
\EE

The \btext{KF is} initialised according to an \emph{a priori} distribution that represents our prior 
knowledge about the system and fixes $\est{\vB{x}}_{0|0}\sim\cN(\vB{\boldsymbol{\mu}}_0,\mB{\Sigma}_0)$. \btext{For} time-invariant system and measurement dynamics \cite{Jazwinski1970}, the KF must reach a \emph{steady-state solution} as $k\to\infty$ with all $\cov_{k|k}$, $\mB{K}_k$, $\mB{S}_k$ converging to steady-state values $\covSSu$, $\mB{K}_{\t{ss}}$, $\mB{S}_{\t{ss}}$, respectively \cite{suppmat}.

\PRLsection{Experiment}%
A cylindrical cell, of length \unit{3}{\centi\meter} and diameter \unit{1}{\centi\meter}, contains isotopically enriched $^{\rm 87}$Rb vapor and 100 Torr of $\rm N_2$ buffer gas, with controlled temperature and magnetic environment \cite{Lucivero2016a} to maintain alkali number density of \unit{4.5 \times 10^{12}}{\centi\meter^{-3}} and $\omega_L = 2\pi \times  \unit{10}{\kilo \hertz}$. Meter light from a distributed-Bragg reflector laser (DBR) is red-detuned by 60 GHz from the  $\rm D_1$ absorption line, while a circularly polarized signal beam from a second DBR diode is tuned to the $\rm D_2$ line-edge. Signal and meter beams each have effective area of  $\unit{0.016}{\centi \meter \squared} $, overlap at a non-polarizing 50:50 beam splitter placed before the cell, and propagate along the $z$ axis. A dichroic high-pass optical filter, placed after the cell, blocks the transmitted $\rm D_2$ light while passing the $\rm D_1$ probe beam for polarization analysis. 
Sensor parameters $\{T_2, Q_y,Q_z, R  \}$ are found by spin noise spectroscopy \cite{Lucivero2016a,Lucivero2016b,suppmat}. The target signal, in particular, its quadratures $\vB{q}_t$, is digitally synthesized using an arbitrary-waveform-generator and applied to the injection current of the signal-beam DBR diode, to produce a pumping rate $\cE(t)$ \cite{Jimenez2010a}. 
\begin{figure}[!t]
\centering
\includegraphics[width=\columnwidth]{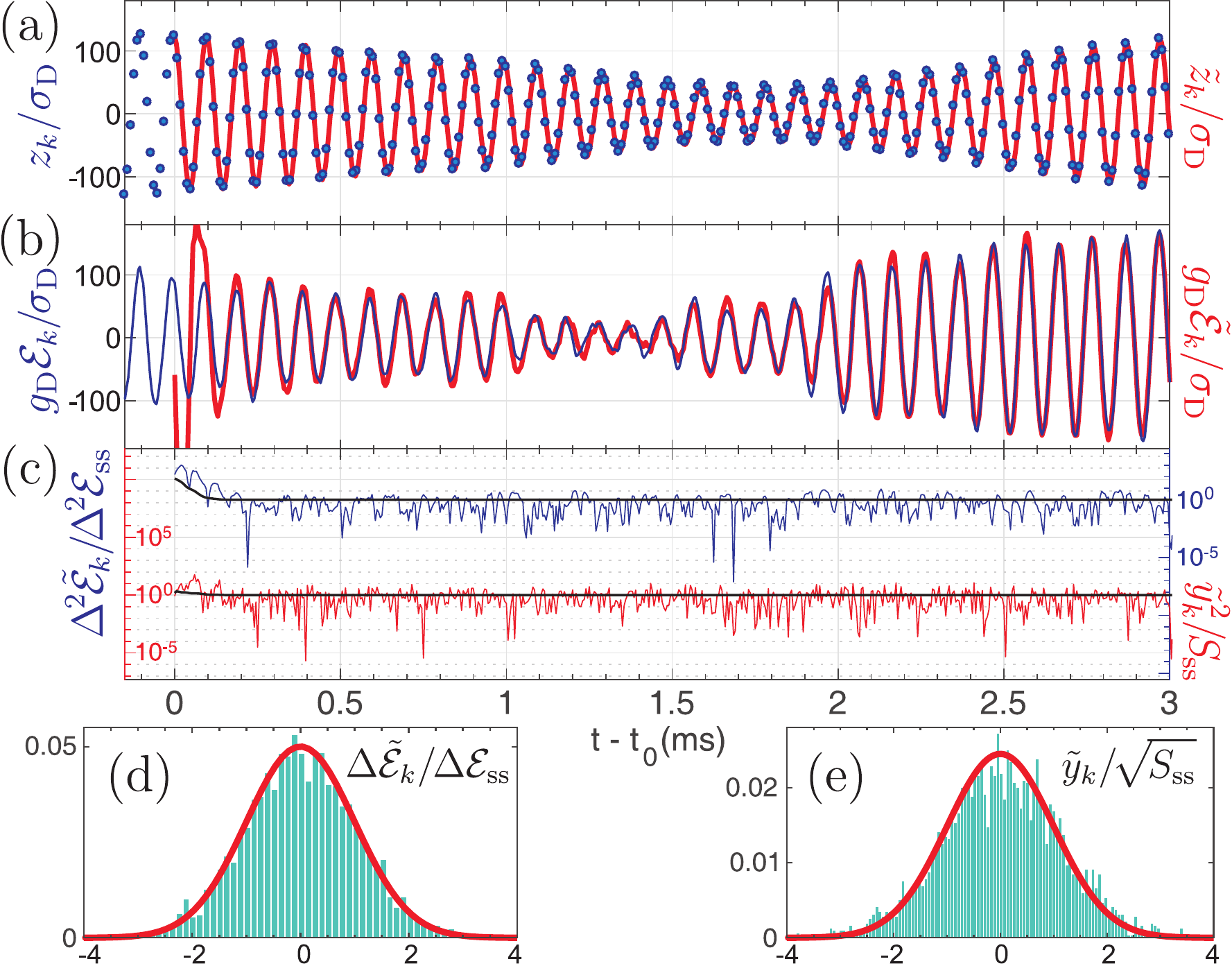}
\caption{(a):~Recorded sensor output ($z_k$, blue dots) along with its KF estimates ($\est{z}_k$, red solid line). The output is sampled at intervals $\Delta =$ \unit{5}{\micro\second};~for clarity only even samples are shown. (b):~Applied amplitude ($\cE(t)$, blue) and its KF estimates ($\est{\cE}_k$, red) shown in optical-rotation angle units scaled to the strength of detection shot-noise
($\sigma_\t{D} = \sqrt{R^\Delta}$). (c):~Corresponding behaviour of 
the \emph{true} waveform estimation error squared ($\Delta^2 \est{\cE}_k$, blue line)
and the innovations squared ($\tilde{y}_k^2$, red line), along with the variances predicted by the KF estimators (black solid lines) and
\eqnsref{Eq:kf-cov-main}{Eq:kf-out-cov}, respectively. (d-e):~Histograms (cyan) of true signal-estimation error and innovations 
collected over a period of $\unit{15}{\milli\second}$---as compared to the Gaussian PDFs predicted by 
the dynamical/observation models employed (red lines). All quantities plotted in (c-e) are renormalised to their asymptotic steady-state solutions.} 
\label{fig:OU-signal}
\end{figure}

\PRLsection{Validation}%
In the first experiment, the waveform 
$\cE(t)$ is a single realization of the OU process described by \eqnref{eq:dyns_pump} with $\kappa= \unit{100}{\second ^{-1}}$, see \figref{fig:setup}(b). A segment of the sensor output sequence, $z_k$, and the applied waveform, $\cE_k$, are shown in \figref{fig:OU-signal}(a-b), along with their KF estimates, $\est{z}_k$ and $\est{\cE}_{k}= \gP(\est{q}_{k|k}\cos(\omega_\t{P} t_k) + \est{p}_{k|k}\sin(\omega_\t{P} t_k))$. In \figref{fig:OU-signal}(c), the square of the corresponding (single-shot) \emph{true} estimation error, $\Delta^2\est{\cE}_{k}=(\est{\cE}_{k}-\cE_{k})^2$,  is plotted along with the (scalar) innovations squared, $\tilde{y}_k^2$. Note that consistently 
with the time-invariant linear-Gaussian model of \eref{eq:dyns_atoms}-\eref{eq:I_meas} the estimates converge to their asymptotic steady-state solutions, i.e., to $\Delta^2\cE_\t{ss}=(\Sigma_\t{ss}^{(q)})^2+(\Sigma_\t{ss}^{(p)})^2$ and $S_\t{ss}$ which we evaluate numerically \cite{suppmat}. Furthermore, in order to fully validate the sensor model and the correctness of the KF implementation, we explicitly verify that the sequences of estimation errors  $\Delta \est{\cE}_k$ and $\est{y}_k$ are described by zero-mean Gaussian processes with variances dictated by \eqnsref{Eq:kf-cov-main}{Eq:kf-out-cov}, respectively \cite{bar2004}. In \figref{fig:OU-signal}(d-e) we compare their histograms with the predicted distributions, and find that 94\% (93\%) of  $\Delta \est{\cE}_k$ ($\est{y}_k$) data points lie within a two-sided 95\% confidence region of their respective predicted Gaussian 
distributions---indicating a very close agreement of the model and observed statistics.
\begin{figure}[t!]
\includegraphics[width=\columnwidth]{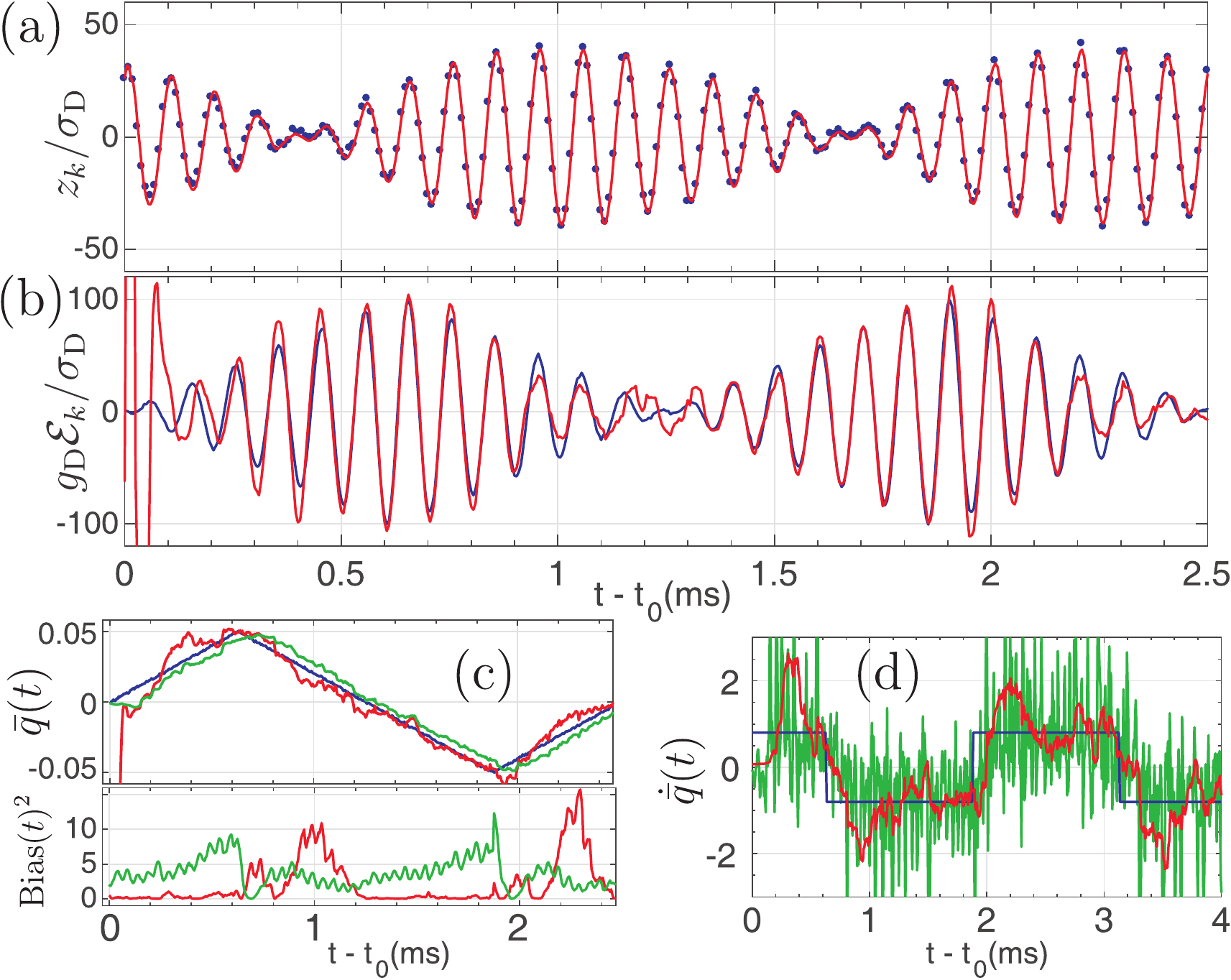}
\caption{%
(a):~Sensor output and;~(b):~applied waveform along with their respective 
KF estimates based on the polynomial model of \eqnref{eq:pm};~colors as in \figref{fig:OU-signal}. (c,~\textit{top}):~ Quadrature $\bar{q}(t)$ of the applied waveform (\textit{blue}), in a single experimental run along with its KF estimates based on Wiener-process (WP, \textit{green}) and polynomial (PM, \textit{red}) models. (c,~\textit{bottom}, same colours):~ Instantaneous bias (squared), $\text{Bias}(t)^{2}$, for the WP and PM estimators (obtained by averaging 6 experimental runs). The consistently delayed response to waveform variations of the WP model (visible comparing the blue and green curves), results in a higher bias of its KF estimate in (c), resulting then (after time-averaging) in a higher MSE shown in \tabref{table:biasandvariance}. (d): Time-derivative of the input waveform, $\dot{\bar{q}}(t)$, with its corresponding WP and PM estimates (same colours). As the WP \btext{estimators do not include} the derivative, $\dot{\bar{q}}(t)$ can only be inferred from $\est{q}_k$, yielding \btext{the} noisy green curve in (d).}
\label{fig:det-signal}
\end{figure}

\PRLsection{Estimating unknown noisy waveforms}%
The ultimate goal of waveform estimation is to track signals with dynamics partially known prior to the measurement \cite{SanderBOE2012,alem2015fetal,bar2004}. We now consider estimating waveforms whose quadrature vector $\vB{q}_t$ follows the quantity $\bar{\vB{q}}_t$ with unknown dynamics, but experiences fluctuations with known statistical properties $\d \vB{w}_t^{(\t q)}\sim\cN(\vB{0},\mB{Q}^{(\t q)}\d t)$, so that:
\BE
\d \vB{q}_t = \dot{\bar{\vB{q}}}_t \d t + \d \vB{w}_t^{(\t q)}. \label{eq:dyns_pump_unknown} 
\EE
To study this scenario we synthesize waveforms with quadrature components $\bar{\vB{q}}(t)=[\bar{q}(t),0]$ and signal noise $\mB{Q}^{(\t q)}=\t{diag}\{Q,Q\}$.\btext{W}e implement our filter with help of a (third-order) \emph{polynomial model} \cite{bar2004,vanTrees2013},, within which the evolution of the $q$-quadrature (and similarly for $p$) is modelled by the following dynamics \cite{suppmat}:
\BE
	\d q(t)             = \dot{q}(t)\d t, ~~
	\d \dot{q}(t)  = \ddot{q}(t)\d t,~~
	\d \ddot{q}(t) =  \d w^{(\ddot{\mathrm{q}})}_t,
\label{eq:pm}
\EE
where the first and second derivatives are treated as a part of the waveform state space. The unknown deterministic and stochastic variations of the signal in \eqnref{eq:dyns_pump_unknown} are then accounted for by introducing effective fluctuations of the $\ddot{q}(t)$-component, $\d w^{(\ddot{\mathrm{q}})}_t =\sqrt{Q/\Delta^4}\,\d W_{\ddot{q}}(t)$, \btext{with variance $\d w^{(\ddot{\mathrm{q}})}_t \d w^{(\ddot{\mathrm{q}})}_t = Q/\Delta^4 \d t$.} As a result, the enlarged quadrature-vector reads:~$\vB{q}_t = [q_t, \dot{q}_t, \ddot{q}_t; p_t, \dot{p}_t, \ddot{p}_t]^T$, and together with the spin degrees of freedom, $\vB{j}_t$, defines now the augmented state space. The KF is then used, as described before, to construct the estimator $\est{\vB{q}}_k$ with help of expressions found in \refcite{suppmat} for the corresponding discrete-time transition- and error-covariance-matrices.

\btext{T}he sensor output sequence $z_k$, applied waveform $\cE_k$, and their respective KF estimates \btext{are shown in \figref{fig:det-signal}.} Similarly to the case of the OU process depicted in \figref{fig:OU-signal}, one observes the output to be distorted (i.e., smoothed and delayed) as compared to the applied waveform. Despite this, the filter tracks the salient features, amplitude and phase, of the waveform in real time. In \mbox{\figref{fig:det-signal}(c-d)}, we show the true evolution of the quadrature $\bar{q}(t)$ and its time-derivative $\dot{\bar{q}}(t)$, respectively, along with the corresponding KF estimates $\est{q}_k$ and $\est{\dot{q}}_k$. We also compare the filter performance against its na\"{i}ve implementation, which assumes the signal to be a pure Wiener process, i.e., $\d \vB{q}_t = \d \vB{w}_t^{(\t q)}$. Although the estimates based on the na\"{i}ve implementation are less noisy, due to a smaller state space, they exhibit an intrinsic delay that cannot be compensated. The precision advantage of the polynomial model is summarised in \tabref{table:biasandvariance}, which shows that despite larger uncertainty (variance) the signal is tracked with much higher precision (smaller MSE) due to significant reduction of the bias. Moreover, by making the signal derivative a part of the state space, $\dot{\bar{q}}(t)$ can now be tracked in real time. In contrast, such information cannot be obtained using the na\"{i}ve implementation of the filter---being masked out by the noise, see \figref{fig:det-signal}(d).  
\begin{table}[t!]
	\centering 
	\begin{tabular}{c| c c c} 
		\hline\hline 
			Model & $\text{Bias}^{2}$  & Var & MSE\\ 
		\hline 	
			$\{\est{q}_k\}_\t{(WP)}$ & ${3.36\times 10^{-5}}{}$ & $ {3.0\times 10^{-6}}{}$   & ${3.66\times 10^{-5}}{}$\\ 
			$\{\est{q}_k\}_\t{(PM)}$ & ${1.02\times 10^{-5}}{}$ & $ {7.6\times 10^{-6}}{}$   & ${1.78\times 10^{-5}}{}$\\			
		\hline 
	\end{tabular}
\caption{%
Squared bias, variance (Var) and mean squared error (MSE)
of the KF estimates, $\{\est{q}_k\}$, for the input quadrature, $\bar{q}(t)$, assuming the signal to be described by the Wiener process (WP) or 
the polynomial model (PM) of \eqnref{eq:pm} and averaging over a time sequence 
of \unit{15}{\milli \second}.}
\label{table:biasandvariance} 
\end{table}

\PRLsection{Conclusions}%
We have \btext{demonstrated} Kalman filtering \btext{in an archetypal} two-stage atomic sensor. Driving the sensor with a known waveform, we have directly confirmed the validity of the statistical model describing the spin dynamics and the optical readout. Incorporating this model into the KF, we have demonstrated the optimal recovery of waveforms with spectral components far outside the intrinsic temporal resolution of the sensor. We have also shown how the same KF techniques can be efficiently employed to track waveforms with dynamics unknown prior to the measurement. These results may pave the way for employing KFs in a wide range of atomic sensing applications \cite{KominisN2003,Geremia2003,KornackPRL2005,AngelisMST2009,Jimenez2014a,SmiciklasPRL2011,ludlow2015, SanderBOE2012,colangelo2017,ciruna2017}.

\begin{acknowledgments}

We thank A. Dimic for the help fabricating magnetic coils and J. B. Brask and M. Tsang for helpful feedback on this work.
This project has received funding from the European Union’s Horizon 2020 research and innovation programme under the Marie Sklodowska-Curie grant agreements QUTEMAG 
(no.~654339) and Q-METAPP (no.~655161). The work was also supported by the European Research Council (ERC) projects AQUMET (280169) 
and ERIDIAN (713682); European Union project QUIC (Grant Agreement no.~641122);~the Spanish MINECO projects MAQRO (Ref. FIS2015-68039-P), XPLICA  (FIS2014-62181-EXP), QIBEQI (Ref. FIS2016-80773-P); the
Severo Ochoa programme (SEV-2015-0522); Ag\`{e}ncia de Gesti\'{o} d'Ajuts Universitaris i de Recerca (AGAUR) project (2014-SGR-1295); Fundaci\'{o} Privada Cellex and Generalitat de Catalunya (CERCA Program).

\end{acknowledgments}

\bibliographystyle{apsrev4-1}
\bibliography{atoms_kalman.bib}

\begin{thebibliography}{40}%
\makeatletter
\providecommand \@ifxundefined [1]{%
 \@ifx{#1\undefined}
}%
\providecommand \@ifnum [1]{%
 \ifnum #1\expandafter \@firstoftwo
 \else \expandafter \@secondoftwo
 \fi
}%
\providecommand \@ifx [1]{%
 \ifx #1\expandafter \@firstoftwo
 \else \expandafter \@secondoftwo
 \fi
}%
\providecommand \natexlab [1]{#1}%
\providecommand \enquote  [1]{``#1''}%
\providecommand \bibnamefont  [1]{#1}%
\providecommand \bibfnamefont [1]{#1}%
\providecommand \citenamefont [1]{#1}%
\providecommand \href@noop [0]{\@secondoftwo}%
\providecommand \href [0]{\begingroup \@sanitize@url \@href}%
\providecommand \@href[1]{\@@startlink{#1}\@@href}%
\providecommand \@@href[1]{\endgroup#1\@@endlink}%
\providecommand \@sanitize@url [0]{\catcode `\\12\catcode `\$12\catcode
  `\&12\catcode `\#12\catcode `\^12\catcode `\_12\catcode `\%12\relax}%
\providecommand \@@startlink[1]{}%
\providecommand \@@endlink[0]{}%
\providecommand \url  [0]{\begingroup\@sanitize@url \@url }%
\providecommand \@url [1]{\endgroup\@href {#1}{\urlprefix }}%
\providecommand \urlprefix  [0]{URL }%
\providecommand \Eprint [0]{\href }%
\providecommand \doibase [0]{http://dx.doi.org/}%
\providecommand \selectlanguage [0]{\@gobble}%
\providecommand \bibinfo  [0]{\@secondoftwo}%
\providecommand \bibfield  [0]{\@secondoftwo}%
\providecommand \translation [1]{[#1]}%
\providecommand \BibitemOpen [0]{}%
\providecommand \bibitemStop [0]{}%
\providecommand \bibitemNoStop [0]{.\EOS\space}%
\providecommand \EOS [0]{\spacefactor3000\relax}%
\providecommand \BibitemShut  [1]{\csname bibitem#1\endcsname}%
\let\auto@bib@innerbib\@empty
\bibitem [{\citenamefont {Ludlow}\ \emph {et~al.}(2015)\citenamefont {Ludlow},
  \citenamefont {Boyd}, \citenamefont {Ye}, \citenamefont {Peik},\ and\
  \citenamefont {Schmidt}}]{ludlow2015}%
  \BibitemOpen
  \bibfield  {author} {\bibinfo {author} {\bibfnamefont {A.~D.}\ \bibnamefont
  {Ludlow}}, \bibinfo {author} {\bibfnamefont {M.~M.}\ \bibnamefont {Boyd}},
  \bibinfo {author} {\bibfnamefont {J.}~\bibnamefont {Ye}}, \bibinfo {author}
  {\bibfnamefont {E.}~\bibnamefont {Peik}}, \ and\ \bibinfo {author}
  {\bibfnamefont {P.~O.}\ \bibnamefont {Schmidt}},\ }\href {\doibase
  10.1103/RevModPhys.87.637} {\bibfield  {journal} {\bibinfo  {journal} {Rev.
  Mod. Phys.}\ }\textbf {\bibinfo {volume} {87}},\ \bibinfo {pages} {637}
  (\bibinfo {year} {2015})}\BibitemShut {NoStop}%
\bibitem [{\citenamefont {Kominis}\ \emph {et~al.}(2003)\citenamefont
  {Kominis}, \citenamefont {Kornack}, \citenamefont {Allred},\ and\
  \citenamefont {Romalis}}]{KominisN2003}%
  \BibitemOpen
  \bibfield  {author} {\bibinfo {author} {\bibfnamefont {I.}~\bibnamefont
  {Kominis}}, \bibinfo {author} {\bibfnamefont {T.}~\bibnamefont {Kornack}},
  \bibinfo {author} {\bibfnamefont {J.}~\bibnamefont {Allred}}, \ and\ \bibinfo
  {author} {\bibfnamefont {M.}~\bibnamefont {Romalis}},\ }\href {\doibase
  10.1038/nature01484} {\bibfield  {journal} {\bibinfo  {journal} {Nature}\
  }\textbf {\bibinfo {volume} {422}},\ \bibinfo {pages} {596} (\bibinfo {year}
  {2003})}\BibitemShut {NoStop}%
\bibitem [{\citenamefont {{LIGO Scientific Collaboration and Virgo
  Collaboration}}(2016)}]{LIGOPRL2016a}%
  \BibitemOpen
  \bibfield  {author} {\bibinfo {author} {\bibnamefont {{LIGO Scientific
  Collaboration and Virgo Collaboration}}},\ }\href {\doibase
  10.1103/PhysRevLett.116.061102} {\bibfield  {journal} {\bibinfo  {journal}
  {Phys. Rev. Lett.}\ }\textbf {\bibinfo {volume} {116}},\ \bibinfo {pages}
  {061102} (\bibinfo {year} {2016})}\BibitemShut {NoStop}%
\bibitem [{\citenamefont {Shah}\ \emph {et~al.}(2010)\citenamefont {Shah},
  \citenamefont {Vasilakis},\ and\ \citenamefont {Romalis}}]{ShahPRL2010}%
  \BibitemOpen
  \bibfield  {author} {\bibinfo {author} {\bibfnamefont {V.}~\bibnamefont
  {Shah}}, \bibinfo {author} {\bibfnamefont {G.}~\bibnamefont {Vasilakis}}, \
  and\ \bibinfo {author} {\bibfnamefont {M.~V.}\ \bibnamefont {Romalis}},\
  }\href {\doibase 10.1103/PhysRevLett.104.013601} {\bibfield  {journal}
  {\bibinfo  {journal} {Phys. Rev. Lett.}\ }\textbf {\bibinfo {volume} {104}},\
  \bibinfo {pages} {013601} (\bibinfo {year} {2010})}\BibitemShut {NoStop}%
\bibitem [{\citenamefont {Sander}\ \emph {et~al.}(2012)\citenamefont {Sander},
  \citenamefont {Preusser}, \citenamefont {Mhaskar}, \citenamefont {Kitching},
  \citenamefont {Trahms},\ and\ \citenamefont {Knappe}}]{SanderBOE2012}%
  \BibitemOpen
  \bibfield  {author} {\bibinfo {author} {\bibfnamefont {T.~H.}\ \bibnamefont
  {Sander}}, \bibinfo {author} {\bibfnamefont {J.}~\bibnamefont {Preusser}},
  \bibinfo {author} {\bibfnamefont {R.}~\bibnamefont {Mhaskar}}, \bibinfo
  {author} {\bibfnamefont {J.}~\bibnamefont {Kitching}}, \bibinfo {author}
  {\bibfnamefont {L.}~\bibnamefont {Trahms}}, \ and\ \bibinfo {author}
  {\bibfnamefont {S.}~\bibnamefont {Knappe}},\ }\href {\doibase
  10.1364/BOE.3.000981} {\bibfield  {journal} {\bibinfo  {journal} {Biomed.
  Opt. Express}\ }\textbf {\bibinfo {volume} {3}},\ \bibinfo {pages} {981}
  (\bibinfo {year} {2012})}\BibitemShut {NoStop}%
\bibitem [{\citenamefont {Tsang}(2009)}]{Tsang2009}%
  \BibitemOpen
  \bibfield  {author} {\bibinfo {author} {\bibfnamefont {M.}~\bibnamefont
  {Tsang}},\ }\href {\doibase 10.1103/PhysRevA.80.033840} {\bibfield  {journal}
  {\bibinfo  {journal} {Phys. Rev. A}\ }\textbf {\bibinfo {volume} {80}},\
  \bibinfo {pages} {033840} (\bibinfo {year} {2009})}\BibitemShut {NoStop}%
\bibitem [{\citenamefont {Tsang}\ \emph {et~al.}(2011)\citenamefont {Tsang},
  \citenamefont {Wiseman},\ and\ \citenamefont {Caves}}]{TsangPRL2011}%
  \BibitemOpen
  \bibfield  {author} {\bibinfo {author} {\bibfnamefont {M.}~\bibnamefont
  {Tsang}}, \bibinfo {author} {\bibfnamefont {H.~M.}\ \bibnamefont {Wiseman}},
  \ and\ \bibinfo {author} {\bibfnamefont {C.~M.}\ \bibnamefont {Caves}},\
  }\href {\doibase 10.1103/PhysRevLett.106.090401} {\bibfield  {journal}
  {\bibinfo  {journal} {Phys. Rev. Lett.}\ }\textbf {\bibinfo {volume} {106}},\
  \bibinfo {pages} {090401} (\bibinfo {year} {2011})}\BibitemShut {NoStop}%
\bibitem [{\citenamefont {Geremia}\ \emph {et~al.}(2004)\citenamefont
  {Geremia}, \citenamefont {Stockton},\ and\ \citenamefont
  {Mabuchi}}]{GeremiaS2004}%
  \BibitemOpen
  \bibfield  {author} {\bibinfo {author} {\bibfnamefont {J.}~\bibnamefont
  {Geremia}}, \bibinfo {author} {\bibfnamefont {J.~K.}\ \bibnamefont
  {Stockton}}, \ and\ \bibinfo {author} {\bibfnamefont {H.}~\bibnamefont
  {Mabuchi}},\ }\href {\doibase 10.1126/science.1095374} {\bibfield  {journal}
  {\bibinfo  {journal} {Science}\ }\textbf {\bibinfo {volume} {304}},\ \bibinfo
  {pages} {270} (\bibinfo {year} {2004})}\BibitemShut {NoStop}%
\bibitem [{\citenamefont {Yonezawa}\ \emph {et~al.}(2012)\citenamefont
  {Yonezawa}, \citenamefont {Nakane}, \citenamefont {Wheatley}, \citenamefont
  {Iwasawa}, \citenamefont {Takeda}, \citenamefont {Arao}, \citenamefont
  {Ohki}, \citenamefont {Tsumura}, \citenamefont {Berry}, \citenamefont
  {Ralph}, \citenamefont {Wiseman}, \citenamefont {Huntington},\ and\
  \citenamefont {Furusawa}}]{Yonezawa2012}%
  \BibitemOpen
  \bibfield  {author} {\bibinfo {author} {\bibfnamefont {H.}~\bibnamefont
  {Yonezawa}}, \bibinfo {author} {\bibfnamefont {D.}~\bibnamefont {Nakane}},
  \bibinfo {author} {\bibfnamefont {T.~A.}\ \bibnamefont {Wheatley}}, \bibinfo
  {author} {\bibfnamefont {K.}~\bibnamefont {Iwasawa}}, \bibinfo {author}
  {\bibfnamefont {S.}~\bibnamefont {Takeda}}, \bibinfo {author} {\bibfnamefont
  {H.}~\bibnamefont {Arao}}, \bibinfo {author} {\bibfnamefont {K.}~\bibnamefont
  {Ohki}}, \bibinfo {author} {\bibfnamefont {K.}~\bibnamefont {Tsumura}},
  \bibinfo {author} {\bibfnamefont {D.~W.}\ \bibnamefont {Berry}}, \bibinfo
  {author} {\bibfnamefont {T.~C.}\ \bibnamefont {Ralph}}, \bibinfo {author}
  {\bibfnamefont {H.~M.}\ \bibnamefont {Wiseman}}, \bibinfo {author}
  {\bibfnamefont {E.~H.}\ \bibnamefont {Huntington}}, \ and\ \bibinfo {author}
  {\bibfnamefont {A.}~\bibnamefont {Furusawa}},\ }\href {\doibase
  10.1126/science.1225258} {\bibfield  {journal} {\bibinfo  {journal}
  {Science}\ }\textbf {\bibinfo {volume} {337}},\ \bibinfo {pages} {1514}
  (\bibinfo {year} {2012})}\BibitemShut {NoStop}%
\bibitem [{\citenamefont {Wieczorek}\ \emph {et~al.}(2015)\citenamefont
  {Wieczorek}, \citenamefont {Hofer}, \citenamefont {Hoelscher-Obermaier},
  \citenamefont {Riedinger}, \citenamefont {Hammerer},\ and\ \citenamefont
  {Aspelmeyer}}]{Wieczorek2015}%
  \BibitemOpen
  \bibfield  {author} {\bibinfo {author} {\bibfnamefont {W.}~\bibnamefont
  {Wieczorek}}, \bibinfo {author} {\bibfnamefont {S.~G.}\ \bibnamefont
  {Hofer}}, \bibinfo {author} {\bibfnamefont {J.}~\bibnamefont
  {Hoelscher-Obermaier}}, \bibinfo {author} {\bibfnamefont {R.}~\bibnamefont
  {Riedinger}}, \bibinfo {author} {\bibfnamefont {K.}~\bibnamefont {Hammerer}},
  \ and\ \bibinfo {author} {\bibfnamefont {M.}~\bibnamefont {Aspelmeyer}},\
  }\href {\doibase 10.1103/PhysRevLett.114.223601} {\bibfield  {journal}
  {\bibinfo  {journal} {Phys. Rev. Lett.}\ }\textbf {\bibinfo {volume} {114}},\
  \bibinfo {pages} {223601} (\bibinfo {year} {2015})}\BibitemShut {NoStop}%
\bibitem [{\citenamefont {Hochberg}\ \emph {et~al.}(2012)\citenamefont
  {Hochberg}, \citenamefont {Bacher}, \citenamefont {Jarosiewicz},
  \citenamefont {Masse}, \citenamefont {Simeral}, \citenamefont {Vogel},
  \citenamefont {Haddadin}, \citenamefont {Liu}, \citenamefont {Cash},
  \citenamefont {van~der Smagt} \emph {et~al.}}]{Hochberg2012}%
  \BibitemOpen
  \bibfield  {author} {\bibinfo {author} {\bibfnamefont {L.~R.}\ \bibnamefont
  {Hochberg}}, \bibinfo {author} {\bibfnamefont {D.}~\bibnamefont {Bacher}},
  \bibinfo {author} {\bibfnamefont {B.}~\bibnamefont {Jarosiewicz}}, \bibinfo
  {author} {\bibfnamefont {N.~Y.}\ \bibnamefont {Masse}}, \bibinfo {author}
  {\bibfnamefont {J.~D.}\ \bibnamefont {Simeral}}, \bibinfo {author}
  {\bibfnamefont {J.}~\bibnamefont {Vogel}}, \bibinfo {author} {\bibfnamefont
  {S.}~\bibnamefont {Haddadin}}, \bibinfo {author} {\bibfnamefont
  {J.}~\bibnamefont {Liu}}, \bibinfo {author} {\bibfnamefont {S.~S.}\
  \bibnamefont {Cash}}, \bibinfo {author} {\bibfnamefont {P.}~\bibnamefont
  {van~der Smagt}},  \emph {et~al.},\ }\href {\doibase 10.1038/nature11076}
  {\bibfield  {journal} {\bibinfo  {journal} {Nature}\ }\textbf {\bibinfo
  {volume} {485}},\ \bibinfo {pages} {372} (\bibinfo {year}
  {2012})}\BibitemShut {NoStop}%
\bibitem [{\citenamefont {Lien}\ \emph {et~al.}(2016)\citenamefont {Lien},
  \citenamefont {Gillian}, \citenamefont {Karagozler}, \citenamefont {Amihood},
  \citenamefont {Schwesig}, \citenamefont {Olson}, \citenamefont {Raja},\ and\
  \citenamefont {Poupyrev}}]{Lien2016}%
  \BibitemOpen
  \bibfield  {author} {\bibinfo {author} {\bibfnamefont {J.}~\bibnamefont
  {Lien}}, \bibinfo {author} {\bibfnamefont {N.}~\bibnamefont {Gillian}},
  \bibinfo {author} {\bibfnamefont {M.~E.}\ \bibnamefont {Karagozler}},
  \bibinfo {author} {\bibfnamefont {P.}~\bibnamefont {Amihood}}, \bibinfo
  {author} {\bibfnamefont {C.}~\bibnamefont {Schwesig}}, \bibinfo {author}
  {\bibfnamefont {E.}~\bibnamefont {Olson}}, \bibinfo {author} {\bibfnamefont
  {H.}~\bibnamefont {Raja}}, \ and\ \bibinfo {author} {\bibfnamefont
  {I.}~\bibnamefont {Poupyrev}},\ }\href {\doibase 10.1145/2897824.2925953}
  {\bibfield  {journal} {\bibinfo  {journal} {ACM Trans. Graph.}\ }\textbf
  {\bibinfo {volume} {35}},\ \bibinfo {pages} {142:1} (\bibinfo {year}
  {2016})}\BibitemShut {NoStop}%
\bibitem [{\citenamefont {van Trees}\ \emph {et~al.}(2013)\citenamefont {van
  Trees}, \citenamefont {Bell},\ and\ \citenamefont {Tian}}]{vanTrees2013}%
  \BibitemOpen
  \bibfield  {author} {\bibinfo {author} {\bibfnamefont {H.~L.}\ \bibnamefont
  {van Trees}}, \bibinfo {author} {\bibfnamefont {K.~L.}\ \bibnamefont {Bell}},
  \ and\ \bibinfo {author} {\bibfnamefont {Z.}~\bibnamefont {Tian}},\
  }\href@noop {} {\emph {\bibinfo {title} {{Detection, Estimation, and
  Modulation Theory. Part I:~Detection, Estimation and Filtering Theory}}}}\
  (\bibinfo  {publisher} {Wiley},\ \bibinfo {year} {2013})\BibitemShut
  {NoStop}%
\bibitem [{\citenamefont {van Trees}\ and\ \citenamefont
  {Bell}(2007)}]{vanTrees2007bayesian}%
  \BibitemOpen
  \bibinfo {editor} {\bibfnamefont {H.~L.}\ \bibnamefont {van Trees}}\ and\
  \bibinfo {editor} {\bibfnamefont {K.~L.}\ \bibnamefont {Bell}},\ eds.,\
  \href@noop {} {\emph {\bibinfo {title} {{Bayesian Bounds for Parameter
  Estimation and Nonlinear Filtering/Tracking}}}}\ (\bibinfo  {publisher}
  {Wiley},\ \bibinfo {year} {2007})\BibitemShut {NoStop}%
\bibitem [{\citenamefont {Kalman}(1960)}]{Kalman1960}%
  \BibitemOpen
  \bibfield  {author} {\bibinfo {author} {\bibfnamefont {R.~E.}\ \bibnamefont
  {Kalman}},\ }\href {\doibase 10.1115/1.3662552} {\bibfield  {journal}
  {\bibinfo  {journal} {J. Basic Eng}\ }\textbf {\bibinfo {volume} {82}},\
  \bibinfo {pages} {35} (\bibinfo {year} {1960})}\BibitemShut {NoStop}%
\bibitem [{\citenamefont {Kalman}\ and\ \citenamefont
  {Bucy}(1961)}]{Kalman1961}%
  \BibitemOpen
  \bibfield  {author} {\bibinfo {author} {\bibfnamefont {R.~E.}\ \bibnamefont
  {Kalman}}\ and\ \bibinfo {author} {\bibfnamefont {R.~S.}\ \bibnamefont
  {Bucy}},\ }\href {\doibase 10.1115/1.3658902} {\bibfield  {journal} {\bibinfo
   {journal} {J. Basic Eng.}\ }\textbf {\bibinfo {volume} {83}},\ \bibinfo
  {pages} {95} (\bibinfo {year} {1961})}\BibitemShut {NoStop}%
\bibitem [{\citenamefont {Durrant-Whyte}\ and\ \citenamefont
  {Bailey}(2006)}]{slam2006}%
  \BibitemOpen
  \bibfield  {author} {\bibinfo {author} {\bibfnamefont {H.}~\bibnamefont
  {Durrant-Whyte}}\ and\ \bibinfo {author} {\bibfnamefont {T.}~\bibnamefont
  {Bailey}},\ }\href {\doibase 10.1109/MRA.2006.1638022} {\bibfield  {journal}
  {\bibinfo  {journal} {IEEE Robot. Autom. Mag.}\ }\textbf {\bibinfo {volume}
  {13}},\ \bibinfo {pages} {99} (\bibinfo {year} {2006})}\BibitemShut {NoStop}%
\bibitem [{\citenamefont {Cassola}\ and\ \citenamefont
  {Burlando}(2012)}]{Cassola2012}%
  \BibitemOpen
  \bibfield  {author} {\bibinfo {author} {\bibfnamefont {F.}~\bibnamefont
  {Cassola}}\ and\ \bibinfo {author} {\bibfnamefont {M.}~\bibnamefont
  {Burlando}},\ }\href {\doibase 10.1016/j.apenergy.2012.03.054} {\bibfield
  {journal} {\bibinfo  {journal} {Appl. Energy}\ }\textbf {\bibinfo {volume}
  {99}},\ \bibinfo {pages} {154} (\bibinfo {year} {2012})}\BibitemShut
  {NoStop}%
\bibitem [{\citenamefont {Iwasawa}\ \emph {et~al.}(2013)\citenamefont
  {Iwasawa}, \citenamefont {Makino}, \citenamefont {Yonezawa}, \citenamefont
  {Tsang}, \citenamefont {Davidovic}, \citenamefont {Huntington},\ and\
  \citenamefont {Furusawa}}]{Iwasawa2013}%
  \BibitemOpen
  \bibfield  {author} {\bibinfo {author} {\bibfnamefont {K.}~\bibnamefont
  {Iwasawa}}, \bibinfo {author} {\bibfnamefont {K.}~\bibnamefont {Makino}},
  \bibinfo {author} {\bibfnamefont {H.}~\bibnamefont {Yonezawa}}, \bibinfo
  {author} {\bibfnamefont {M.}~\bibnamefont {Tsang}}, \bibinfo {author}
  {\bibfnamefont {A.}~\bibnamefont {Davidovic}}, \bibinfo {author}
  {\bibfnamefont {E.}~\bibnamefont {Huntington}}, \ and\ \bibinfo {author}
  {\bibfnamefont {A.}~\bibnamefont {Furusawa}},\ }\href {\doibase
  10.1103/PhysRevLett.111.163602} {\bibfield  {journal} {\bibinfo  {journal}
  {Phys. Rev. Lett.}\ }\textbf {\bibinfo {volume} {111}},\ \bibinfo {pages}
  {163602} (\bibinfo {year} {2013})}\BibitemShut {NoStop}%
\bibitem [{\citenamefont {Lucivero}\ \emph {et~al.}(2014)\citenamefont
  {Lucivero}, \citenamefont {Anielski}, \citenamefont {Gawlik},\ and\
  \citenamefont {Mitchell}}]{Lucivero2014}%
  \BibitemOpen
  \bibfield  {author} {\bibinfo {author} {\bibfnamefont {V.~G.}\ \bibnamefont
  {Lucivero}}, \bibinfo {author} {\bibfnamefont {P.}~\bibnamefont {Anielski}},
  \bibinfo {author} {\bibfnamefont {W.}~\bibnamefont {Gawlik}}, \ and\ \bibinfo
  {author} {\bibfnamefont {M.~W.}\ \bibnamefont {Mitchell}},\ }\href {\doibase
  10.1063/1.4901588} {\bibfield  {journal} {\bibinfo  {journal} {Rev. Sci.
  Ins.}\ }\textbf {\bibinfo {volume} {85}},\ \bibinfo {pages} {113108}
  (\bibinfo {year} {2014})}\BibitemShut {NoStop}%
\bibitem [{\citenamefont {Kornack}\ \emph {et~al.}(2005)\citenamefont
  {Kornack}, \citenamefont {Ghosh},\ and\ \citenamefont
  {Romalis}}]{KornackPRL2005}%
  \BibitemOpen
  \bibfield  {author} {\bibinfo {author} {\bibfnamefont {T.~W.}\ \bibnamefont
  {Kornack}}, \bibinfo {author} {\bibfnamefont {R.~K.}\ \bibnamefont {Ghosh}},
  \ and\ \bibinfo {author} {\bibfnamefont {M.~V.}\ \bibnamefont {Romalis}},\
  }\href {\doibase 10.1103/PhysRevLett.95.230801} {\bibfield  {journal}
  {\bibinfo  {journal} {Phys. Rev. Lett.}\ }\textbf {\bibinfo {volume} {95}},\
  \bibinfo {pages} {230801} (\bibinfo {year} {2005})}\BibitemShut {NoStop}%
\bibitem [{\citenamefont {de~Angelis}\ \emph {et~al.}(2009)\citenamefont
  {de~Angelis}, \citenamefont {Bertoldi}, \citenamefont {Cacciapuoti},
  \citenamefont {Giorgini}, \citenamefont {Lamporesi}, \citenamefont
  {Prevedelli}, \citenamefont {Saccorotti}, \citenamefont {Sorrentino},\ and\
  \citenamefont {Tino}}]{AngelisMST2009}%
  \BibitemOpen
  \bibfield  {author} {\bibinfo {author} {\bibfnamefont {M.}~\bibnamefont
  {de~Angelis}}, \bibinfo {author} {\bibfnamefont {A.}~\bibnamefont
  {Bertoldi}}, \bibinfo {author} {\bibfnamefont {L.}~\bibnamefont
  {Cacciapuoti}}, \bibinfo {author} {\bibfnamefont {A.}~\bibnamefont
  {Giorgini}}, \bibinfo {author} {\bibfnamefont {G.}~\bibnamefont {Lamporesi}},
  \bibinfo {author} {\bibfnamefont {M.}~\bibnamefont {Prevedelli}}, \bibinfo
  {author} {\bibfnamefont {G.}~\bibnamefont {Saccorotti}}, \bibinfo {author}
  {\bibfnamefont {F.}~\bibnamefont {Sorrentino}}, \ and\ \bibinfo {author}
  {\bibfnamefont {G.~M.}\ \bibnamefont {Tino}},\ }\href {\doibase
  {10.1088/0957-0233/20/2/022001}} {\bibfield  {journal} {\bibinfo  {journal}
  {{Meas. Sci. Technol.}}\ }\textbf {\bibinfo {volume} {{20}}},\ \bibinfo
  {pages} {{022001}} (\bibinfo {year} {{2009}})}\BibitemShut {NoStop}%
\bibitem [{\citenamefont {Jim{\'e}nez-Mart{\'\i}nez}\ \emph
  {et~al.}(2014)\citenamefont {Jim{\'e}nez-Mart{\'\i}nez}, \citenamefont
  {Kennedy}, \citenamefont {Rosenbluh}, \citenamefont {Donley}, \citenamefont
  {Knappe}, \citenamefont {Seltzer}, \citenamefont {Ring}, \citenamefont
  {Bajaj},\ and\ \citenamefont {Kitching}}]{Jimenez2014a}%
  \BibitemOpen
  \bibfield  {author} {\bibinfo {author} {\bibfnamefont {R.}~\bibnamefont
  {Jim{\'e}nez-Mart{\'\i}nez}}, \bibinfo {author} {\bibfnamefont {D.~J.}\
  \bibnamefont {Kennedy}}, \bibinfo {author} {\bibfnamefont {M.}~\bibnamefont
  {Rosenbluh}}, \bibinfo {author} {\bibfnamefont {E.~A.}\ \bibnamefont
  {Donley}}, \bibinfo {author} {\bibfnamefont {S.}~\bibnamefont {Knappe}},
  \bibinfo {author} {\bibfnamefont {S.~J.}\ \bibnamefont {Seltzer}}, \bibinfo
  {author} {\bibfnamefont {H.~L.}\ \bibnamefont {Ring}}, \bibinfo {author}
  {\bibfnamefont {V.~S.}\ \bibnamefont {Bajaj}}, \ and\ \bibinfo {author}
  {\bibfnamefont {J.}~\bibnamefont {Kitching}},\ }\href
  {http://dx.doi.org/10.1038/ncomms4908} {\bibfield  {journal} {\bibinfo
  {journal} {Nat. Commun.}\ }\textbf {\bibinfo {volume} {5}},\ \bibinfo {pages}
  {3908 EP } (\bibinfo {year} {2014})}\BibitemShut {NoStop}%
\bibitem [{\citenamefont {Smiciklas}\ \emph {et~al.}(2011)\citenamefont
  {Smiciklas}, \citenamefont {Brown}, \citenamefont {Cheuk}, \citenamefont
  {Smullin},\ and\ \citenamefont {Romalis}}]{SmiciklasPRL2011}%
  \BibitemOpen
  \bibfield  {author} {\bibinfo {author} {\bibfnamefont {M.}~\bibnamefont
  {Smiciklas}}, \bibinfo {author} {\bibfnamefont {J.~M.}\ \bibnamefont
  {Brown}}, \bibinfo {author} {\bibfnamefont {L.~W.}\ \bibnamefont {Cheuk}},
  \bibinfo {author} {\bibfnamefont {S.~J.}\ \bibnamefont {Smullin}}, \ and\
  \bibinfo {author} {\bibfnamefont {M.~V.}\ \bibnamefont {Romalis}},\ }\href
  {\doibase 10.1103/PhysRevLett.107.171604} {\bibfield  {journal} {\bibinfo
  {journal} {Phys. Rev. Lett.}\ }\textbf {\bibinfo {volume} {107}},\ \bibinfo
  {pages} {171604} (\bibinfo {year} {2011})}\BibitemShut {NoStop}%
\bibitem [{\citenamefont {Julsgaard}\ \emph {et~al.}(2004)\citenamefont
  {Julsgaard}, \citenamefont {Sherson}, \citenamefont {Cirac}, \citenamefont
  {Fiur{\'a}{\v{s}}ek},\ and\ \citenamefont {Polzik}}]{julsgaard2004}%
  \BibitemOpen
  \bibfield  {author} {\bibinfo {author} {\bibfnamefont {B.}~\bibnamefont
  {Julsgaard}}, \bibinfo {author} {\bibfnamefont {J.}~\bibnamefont {Sherson}},
  \bibinfo {author} {\bibfnamefont {J.~I.}\ \bibnamefont {Cirac}}, \bibinfo
  {author} {\bibfnamefont {J.}~\bibnamefont {Fiur{\'a}{\v{s}}ek}}, \ and\
  \bibinfo {author} {\bibfnamefont {E.~S.}\ \bibnamefont {Polzik}},\ }\href
  {\doibase 10.1038/nature03064} {\bibfield  {journal} {\bibinfo  {journal}
  {Nature}\ }\textbf {\bibinfo {volume} {432}},\ \bibinfo {pages} {482}
  (\bibinfo {year} {2004})}\BibitemShut {NoStop}%
\bibitem [{\citenamefont {Alem}\ \emph {et~al.}(2015)\citenamefont {Alem},
  \citenamefont {Sander}, \citenamefont {Mhaskar}, \citenamefont {LeBlanc},
  \citenamefont {Eswaran}, \citenamefont {Steinhoff}, \citenamefont {Okada},
  \citenamefont {Kitching}, \citenamefont {Trahms},\ and\ \citenamefont
  {Knappe}}]{alem2015fetal}%
  \BibitemOpen
  \bibfield  {author} {\bibinfo {author} {\bibfnamefont {O.}~\bibnamefont
  {Alem}}, \bibinfo {author} {\bibfnamefont {T.~H.}\ \bibnamefont {Sander}},
  \bibinfo {author} {\bibfnamefont {R.}~\bibnamefont {Mhaskar}}, \bibinfo
  {author} {\bibfnamefont {J.}~\bibnamefont {LeBlanc}}, \bibinfo {author}
  {\bibfnamefont {H.}~\bibnamefont {Eswaran}}, \bibinfo {author} {\bibfnamefont
  {U.}~\bibnamefont {Steinhoff}}, \bibinfo {author} {\bibfnamefont
  {Y.}~\bibnamefont {Okada}}, \bibinfo {author} {\bibfnamefont
  {J.}~\bibnamefont {Kitching}}, \bibinfo {author} {\bibfnamefont
  {L.}~\bibnamefont {Trahms}}, \ and\ \bibinfo {author} {\bibfnamefont
  {S.}~\bibnamefont {Knappe}},\ }\href {\doibase 10.1088/0031-9155/60/12/4797}
  {\bibfield  {journal} {\bibinfo  {journal} {Phys. Med. Biol.}\ }\textbf
  {\bibinfo {volume} {60}},\ \bibinfo {pages} {4797} (\bibinfo {year}
  {2015})}\BibitemShut {NoStop}%
\bibitem [{\citenamefont {Bar-Shalom}\ \emph {et~al.}(2004)\citenamefont
  {Bar-Shalom}, \citenamefont {Li},\ and\ \citenamefont
  {Kirubarajan}}]{bar2004}%
  \BibitemOpen
  \bibfield  {author} {\bibinfo {author} {\bibfnamefont {Y.}~\bibnamefont
  {Bar-Shalom}}, \bibinfo {author} {\bibfnamefont {X.~R.}\ \bibnamefont {Li}},
  \ and\ \bibinfo {author} {\bibfnamefont {T.}~\bibnamefont {Kirubarajan}},\
  }\href@noop {} {\emph {\bibinfo {title} {{Estimation with Applications to
  Tracking and Navigation: Theory Algorithms and Software}}}}\ (\bibinfo
  {publisher} {Wiley},\ \bibinfo {year} {2004})\BibitemShut {NoStop}%
\bibitem [{\citenamefont {Lucivero}\ \emph {et~al.}(2016)\citenamefont
  {Lucivero}, \citenamefont {Jim\'enez-Mart\'{\i}nez}, \citenamefont {Kong},\
  and\ \citenamefont {Mitchell}}]{Lucivero2016a}%
  \BibitemOpen
  \bibfield  {author} {\bibinfo {author} {\bibfnamefont {V.~G.}\ \bibnamefont
  {Lucivero}}, \bibinfo {author} {\bibfnamefont {R.}~\bibnamefont
  {Jim\'enez-Mart\'{\i}nez}}, \bibinfo {author} {\bibfnamefont
  {J.}~\bibnamefont {Kong}}, \ and\ \bibinfo {author} {\bibfnamefont {M.~W.}\
  \bibnamefont {Mitchell}},\ }\href {\doibase 10.1103/PhysRevA.93.053802}
  {\bibfield  {journal} {\bibinfo  {journal} {Phys. Rev. A}\ }\textbf {\bibinfo
  {volume} {93}},\ \bibinfo {pages} {053802} (\bibinfo {year}
  {2016})}\BibitemShut {NoStop}%
\bibitem [{\citenamefont {Lucivero}\ \emph {et~al.}(2017)\citenamefont
  {Lucivero}, \citenamefont {Dimic}, \citenamefont {Kong}, \citenamefont
  {Jim\'enez-Mart\'{\i}nez},\ and\ \citenamefont {Mitchell}}]{Lucivero2016b}%
  \BibitemOpen
  \bibfield  {author} {\bibinfo {author} {\bibfnamefont {V.~G.}\ \bibnamefont
  {Lucivero}}, \bibinfo {author} {\bibfnamefont {A.}~\bibnamefont {Dimic}},
  \bibinfo {author} {\bibfnamefont {J.}~\bibnamefont {Kong}}, \bibinfo {author}
  {\bibfnamefont {R.}~\bibnamefont {Jim\'enez-Mart\'{\i}nez}}, \ and\ \bibinfo
  {author} {\bibfnamefont {M.~W.}\ \bibnamefont {Mitchell}},\ }\href {\doibase
  10.1103/PhysRevA.95.041803} {\bibfield  {journal} {\bibinfo  {journal} {Phys.
  Rev. A}\ }\textbf {\bibinfo {volume} {95}},\ \bibinfo {pages} {041803}
  (\bibinfo {year} {2017})}\BibitemShut {NoStop}%
\bibitem [{sup()}]{suppmat}%
  \BibitemOpen
  \href@noop {} {}\bibinfo {note} {See Supplemental Material for a description
  of the {K}alman-{B}ucy filter, details on the implementation of the
  continuous-discrete {K}alman filter in our experiment, and sensor
  characterization.}\BibitemShut {Stop}%
\bibitem [{\citenamefont {Gardiner}(1985)}]{gardiner1985}%
  \BibitemOpen
  \bibfield  {author} {\bibinfo {author} {\bibfnamefont {C.~W.}\ \bibnamefont
  {Gardiner}},\ }\href@noop {} {\emph {\bibinfo {title} {{Handbook of
  Stochastic Methods}}}},\ \bibinfo {edition} {3rd}\ ed.\ (\bibinfo
  {publisher} {Springer},\ \bibinfo {year} {1985})\BibitemShut {NoStop}%
\bibitem [{\citenamefont {Tsang}\ \emph {et~al.}(2009)\citenamefont {Tsang},
  \citenamefont {Shapiro},\ and\ \citenamefont {Lloyd}}]{Tsang2009PRA}%
  \BibitemOpen
  \bibfield  {author} {\bibinfo {author} {\bibfnamefont {M.}~\bibnamefont
  {Tsang}}, \bibinfo {author} {\bibfnamefont {J.~H.}\ \bibnamefont {Shapiro}},
  \ and\ \bibinfo {author} {\bibfnamefont {S.}~\bibnamefont {Lloyd}},\ }\href
  {\doibase 10.1103/PhysRevA.79.053843} {\bibfield  {journal} {\bibinfo
  {journal} {Phys. Rev. A}\ }\textbf {\bibinfo {volume} {79}},\ \bibinfo
  {pages} {053843} (\bibinfo {year} {2009})}\BibitemShut {NoStop}%
\bibitem [{\citenamefont {Jazwinski}(1970)}]{Jazwinski1970}%
  \BibitemOpen
  \bibfield  {author} {\bibinfo {author} {\bibfnamefont {A.~H.}\ \bibnamefont
  {Jazwinski}},\ }\href@noop {} {\emph {\bibinfo {title} {{Stochastic Processes
  and Filtering Theory}}}}\ (\bibinfo  {publisher} {Academic Press},\ \bibinfo
  {year} {1970})\BibitemShut {NoStop}%
\bibitem [{\citenamefont {Jimenez-Martinez}\ \emph {et~al.}(2010)\citenamefont
  {Jimenez-Martinez}, \citenamefont {Griffith}, \citenamefont {Wang},
  \citenamefont {Knappe}, \citenamefont {Kitching}, \citenamefont {Smith},\
  and\ \citenamefont {Prouty}}]{Jimenez2010a}%
  \BibitemOpen
  \bibfield  {author} {\bibinfo {author} {\bibfnamefont {R.}~\bibnamefont
  {Jimenez-Martinez}}, \bibinfo {author} {\bibfnamefont {W.~C.}\ \bibnamefont
  {Griffith}}, \bibinfo {author} {\bibfnamefont {Y.~J.}\ \bibnamefont {Wang}},
  \bibinfo {author} {\bibfnamefont {S.}~\bibnamefont {Knappe}}, \bibinfo
  {author} {\bibfnamefont {J.}~\bibnamefont {Kitching}}, \bibinfo {author}
  {\bibfnamefont {K.}~\bibnamefont {Smith}}, \ and\ \bibinfo {author}
  {\bibfnamefont {M.~D.}\ \bibnamefont {Prouty}},\ }\href {\doibase
  10.1109/TIM.2009.2023829} {\bibfield  {journal} {\bibinfo  {journal} {IEEE
  Trans. Instrum. Meas.}\ }\textbf {\bibinfo {volume} {59}},\ \bibinfo {pages}
  {372} (\bibinfo {year} {2010})}\BibitemShut {NoStop}%
\bibitem [{\citenamefont {Geremia}\ \emph {et~al.}(2003)\citenamefont
  {Geremia}, \citenamefont {Stockton}, \citenamefont {Doherty},\ and\
  \citenamefont {Mabuchi}}]{Geremia2003}%
  \BibitemOpen
  \bibfield  {author} {\bibinfo {author} {\bibfnamefont {J.}~\bibnamefont
  {Geremia}}, \bibinfo {author} {\bibfnamefont {J.~K.}\ \bibnamefont
  {Stockton}}, \bibinfo {author} {\bibfnamefont {A.~C.}\ \bibnamefont
  {Doherty}}, \ and\ \bibinfo {author} {\bibfnamefont {H.}~\bibnamefont
  {Mabuchi}},\ }\href {\doibase 10.1103/PhysRevLett.91.250801} {\bibfield
  {journal} {\bibinfo  {journal} {Phys. Rev. Lett.}\ }\textbf {\bibinfo
  {volume} {91}},\ \bibinfo {pages} {250801} (\bibinfo {year}
  {2003})}\BibitemShut {NoStop}%
\bibitem [{\citenamefont {Colangelo}\ \emph {et~al.}(2017)\citenamefont
  {Colangelo}, \citenamefont {Ciurana}, \citenamefont {Bianchet}, \citenamefont
  {Sewell},\ and\ \citenamefont {Mitchell}}]{colangelo2017}%
  \BibitemOpen
  \bibfield  {author} {\bibinfo {author} {\bibfnamefont {G.}~\bibnamefont
  {Colangelo}}, \bibinfo {author} {\bibfnamefont {F.~M.}\ \bibnamefont
  {Ciurana}}, \bibinfo {author} {\bibfnamefont {L.~C.}\ \bibnamefont
  {Bianchet}}, \bibinfo {author} {\bibfnamefont {R.~J.}\ \bibnamefont
  {Sewell}}, \ and\ \bibinfo {author} {\bibfnamefont {M.~W.}\ \bibnamefont
  {Mitchell}},\ }\href {\doibase 10.1038/nature21434} {\bibfield  {journal}
  {\bibinfo  {journal} {Nature}\ }\textbf {\bibinfo {volume} {543}},\ \bibinfo
  {pages} {525} (\bibinfo {year} {2017})}\BibitemShut {NoStop}%
\bibitem [{\citenamefont {Martin~Ciurana}\ \emph {et~al.}(2017)\citenamefont
  {Martin~Ciurana}, \citenamefont {Colangelo}, \citenamefont
  {Slodi\ifmmode~\check{c}\else \v{c}\fi{}ka}, \citenamefont {Sewell},\ and\
  \citenamefont {Mitchell}}]{ciruna2017}%
  \BibitemOpen
  \bibfield  {author} {\bibinfo {author} {\bibfnamefont {F.}~\bibnamefont
  {Martin~Ciurana}}, \bibinfo {author} {\bibfnamefont {G.}~\bibnamefont
  {Colangelo}}, \bibinfo {author} {\bibfnamefont {L.}~\bibnamefont
  {Slodi\ifmmode~\check{c}\else \v{c}\fi{}ka}}, \bibinfo {author}
  {\bibfnamefont {R.~J.}\ \bibnamefont {Sewell}}, \ and\ \bibinfo {author}
  {\bibfnamefont {M.~W.}\ \bibnamefont {Mitchell}},\ }\href {\doibase
  10.1103/PhysRevLett.119.043603} {\bibfield  {journal} {\bibinfo  {journal}
  {Phys. Rev. Lett.}\ }\textbf {\bibinfo {volume} {119}},\ \bibinfo {pages}
  {043603} (\bibinfo {year} {2017})}\BibitemShut {NoStop}%
\bibitem [{\citenamefont {Happer}(1972)}]{happerRMP1972}%
  \BibitemOpen
  \bibfield  {author} {\bibinfo {author} {\bibfnamefont {W.}~\bibnamefont
  {Happer}},\ }\href {\doibase 10.1103/RevModPhys.44.169} {\bibfield  {journal}
  {\bibinfo  {journal} {Rev. Mod. Phys.}\ }\textbf {\bibinfo {volume} {44}},\
  \bibinfo {pages} {169} (\bibinfo {year} {1972})}\BibitemShut {NoStop}%
\bibitem [{\citenamefont {Biedenharn}\ and\ \citenamefont
  {Louck}(1981)}]{Biedenharn1984}%
  \BibitemOpen
  \bibfield  {author} {\bibinfo {author} {\bibfnamefont {L.~C.}\ \bibnamefont
  {Biedenharn}}\ and\ \bibinfo {author} {\bibfnamefont {J.~D.}\ \bibnamefont
  {Louck}},\ }\href@noop {} {\emph {\bibinfo {title} {Angular Momentum in
  Quantum Physics}}}\ (\bibinfo  {publisher} {Addison Wesley},\ \bibinfo {year}
  {1981})\BibitemShut {NoStop}%
\bibitem [{\citenamefont {Laub}(1979)}]{Laub1979}%
  \BibitemOpen
  \bibfield  {author} {\bibinfo {author} {\bibfnamefont {A.}~\bibnamefont
  {Laub}},\ }\href {\doibase 10.1109/TAC.1979.1102178} {\bibfield  {journal}
  {\bibinfo  {journal} {IEEE Trans. Autom. Control}\ }\textbf {\bibinfo
  {volume} {24}},\ \bibinfo {pages} {913} (\bibinfo {year} {1979})}\BibitemShut
  {NoStop}%
\end{thebibliography}%


\onecolumngrid
\clearpage
\noindent
{\centering {\Large \textbf{Supplementary material for}}\\
\vspace{2mm}
{\large R. Jim\'{e}nez-Mart\'{i}nez et al.~``\textit{\MyTitle}'' \\}}
\appendix
\tableofcontents
\setcounter{tocdepth}{5} 
\setcounter{secnumdepth}{5}
\appendix

\section{Optical detection of atomic spin and sensor characterisation} 
\label{app:detect_sensor_char}

\subsection{Faraday optical-rotation}
Hyperfine coupling between the nuclear, $I$, and electronic, $S=1/2$, spins of an alkali atom splits its ground state into two hyperfine manifolds with total angular momentum:~$f_{a} = I + 1/2 $ and $f_{b} = I - 1/2$ (here, we set $\hbar\!=\!1$) \cite{happerRMP1972}. As a result, the Faraday optical-rotation angle $\Theta_\t{FR}$ experienced by linearly-polarized off-resonance light propagating along the $z$ axis, and interacting with $N$ alkali atoms in the ground state, reads 
\BE
	\Theta_{\rm FR} = \frac{1}{2\INucl +1}\left(g_a F_{a,z} - g_b F_{b,z}\right),
	\label{Ap-eq:for}
\EE
where $F_{\alpha,z}$ corresponds to the expectation value of the $z$-component of the collective spin of probed atoms associated with the hyperfine level $\alpha$, i.e., $F_{\alpha,z} \!=\! \tr{\rho^N \sum_{k=1}^N {\hat f}^{(k)}_{\alpha,z}}$ with $\rho^N$ being the ground-state density matrix describing the probed atomic ensemble, while ${\hat f}^{(k)}_{\alpha,z}$ represents the relevant angular momentum of the $k$th atom. In particular, as in our experiments the atoms are prepared in a coherent-spin state \cite{happerRMP1972} that is separable and permutation invariant, i.e, $\rho^N=\varrho^{\ot N}$, the collective spin operators for any $\alpha$th hyperfine level just linearly add, so that
$F_{\alpha,z}\!=\! N f_{\alpha,z}$ where $f_{\alpha,z}\!=\!\tr{\varrho{\hat f}_{\alpha,z}}$ stands for the mean angular momentum of each individual atom. Here, $N = nA_\t{eff}L$ denotes the number of probed alkali atoms with $n$ being the alkali vapor density, $L$ is the path length of the light beam and $A_{\rm eff}$ its effective area \cite{Lucivero2016a,ShahPRL2010}. 

The hyperfine-coupling constant $g_\alpha$ in \eqnref{Ap-eq:for} is given by \cite{happerRMP1972,ShahPRL2010}
\BE
	g_{\alpha} = \frac{c\,r_\t{e}\,f_\t{osc}}{A_{\rm eff}}\,\frac{\nu-\nu_{\alpha}'}{(\nu-\nu_{\alpha})^2+(\Delta\nu_{\rm D1}/2)^2},
	\label{Apeq:coupling-hf}
\EE
where $r_\t{e} = 2.82 \times 10^{-13} $ cm is the classical electron radius, $f_\t{osc}=0.34$ is the oscillator strength of the $\t{D}_1$ transition in Rb, and $c$ is the speed of light.
In \eqnref{Apeq:coupling-hf}, $\Delta \nu_{\rm D1}/2$ represents the pressure-broadened full-width at half-maximum (FWHM) of the $\t{D}_1$ optical transition and  $\nu - \nu_{\alpha}$ denotes the optical detuning of the probe-light. For our experimental conditions, i.e., alkali vapour cell filled with $100$ Torr of $\mathrm{N}_2$ buffer gas, $\Delta\nulight/2 \approx \unit{2.4}{\giga\hertz}$. For a far-detuned probe-light beam, such that $|\nu-1/2\left(\nu_{a}+\nu_{b}\right)| >> |\nu_{a}-\nu_{b}|$, one can approximate $g_a \!\approx\! g_b$.
Using the Wigner-Eckart theorem one obtains $ j_{z}\!=\!\left(2\INucl +1\right)^{-1} (f_{a,z} - f_{b,z})$
\cite{Biedenharn1984}. Thus, for the far-detuned light beam used in our experiments $\Theta_{\rm FR}$ can be approximated by  
\BE
	\Theta_{\rm FR} \approx  \frac{c\,r_\t{e}\,f_\t{osc}}{A_{\rm eff}}\frac{1}{(\nu-\nu_{\alpha}')} J_z,  	
	\label{eq:angle_FR}
\EE
where $J_z = N j_{z} = NP_z/2 $ denotes the mean value of the collective spin along the $z$ direction, with the hyperfine structure ignored, with $P_z \in [-1,1]$ being the electronic spin polarization.

\subsection{Detector photocurrent}
\label{sec:det_photocurr}
To detect the optical rotation angle $\Theta_{\rm FR}$ of the meter light we use a balanced polarimeter consisting of a half-wave plate, polarization-beam-splitter, two balanced photodiodes and a low-noise transimpedance amplifier (TIA). The output of the TIA is given by $V_{\rm DPD}(t)  =  GI(t)$ with $G =\unit{10^{6}}{ \volt \per {\ampere}}$ being the TIA gain and the photocurrent $I(t)$, which in the limit $\Theta_{\rm FR} \ll 1$ describing our experimental conditions is given by 
\BE
    I(t) \d t  =  2 \Re P \Theta_{\rm FR}(t) \d t  + \d w_{\rm sn}(t) ,
    \label{Eq:VPD}
\EE
where $P =\int_{\cal A} dxdy\, \Intens(x,y)$ is the total power of the probe beam of area $ \cal{A} = \rm A_{\rm eff} = \unit{0.016}{\centi\meter\squared} $ reaching the detector with intensity profile $\Intens(x,y)$ and $\Re = \unit{0.59}{\ampere/\watt}$  corresponds to the photodiode's responsivity. In the first (second) experiment reported in the main manuscript $P = \unit{500}{\micro\watt} $ ($P = \unit{100}{\micro\watt} $). In \eqnref{Eq:VPD}, $ \d w_{\rm sn}(t) =\sqrt{R}\d W$, where $\d W\sim\cN(0,\d t)$ is the differential Wiener increment \cite{gardiner1985} and $R$ represents the intensity of the light shot-noise.  

In our experiments the photocurrent $I(t)$ is sampled at a rate $\Delta^{-1} = \unit{200}{\kilo Sa-\second}$. Thus, in order to correctly interpret the measurement outcomes, we need to formulate a discrete-time version of \eqnref{Eq:VPD}. Viewing the sampling process as a short-term average of the continuous-time measurement (c.f.~\cite{bar2004}) the photocurrent $I(t_k)$ recorded at $t_k\!=\!k\,\Delta$, with $k$ being an integer, can be expressed as
\BE
	I(t_k) = \frac{1}{\Delta} \int _{t_k-\Delta}^{t_k} I(t') \d t' =  2 \Re P \Theta_{\rm FR}(t_k) +  \frac{1}{\Delta} \int _{t_k-\Delta}^{t_k} \d w_{\rm sn}(t').
	\label{Apeq:photo}
\EE
Hence, interpreting the last term above as an effective Langevin noise, i.e.,
\BE
	\xi_\t{D}(t_k) \equiv \frac{1}{\Delta} \int _{t_k-\Delta}^{t_k} \d w_{\rm sn}(t')
\label{eq:discretised_obs_noise}
\EE
such that $R^\Delta := \mean{\xi_\t{D}(t)\xi_\t{D}(t)} = R/\Delta$, with $\Delta^{-1}$ quantifying the effective 
noise-bandwidth of each observation, and substituting for the Faraday optical-rotation angle according to \eqnref{eq:angle_FR}, one 
finally arrives at \eqnref{eq:I_meas} of the main text that describes the discrete-time detection process. The effective coupling constant $\gD$ 
in \eqnref{eq:I_meas}, which describes the (linear) transduction between the atomic spin and the photocurrent mediated by the meter light, then reads
\BE
	\gD = 2\Re P \frac{c\,r_\t{e}\,f_\t{osc}}{A_{\rm eff}}\frac{1}{(\nu-\nu_{\alpha}')}.
	\label{Apeq:coupling}
\EE 
\begin{figure}[!t]
\begin{minipage}[c]{0.5\linewidth}
\centering
\includegraphics[width=\columnwidth]{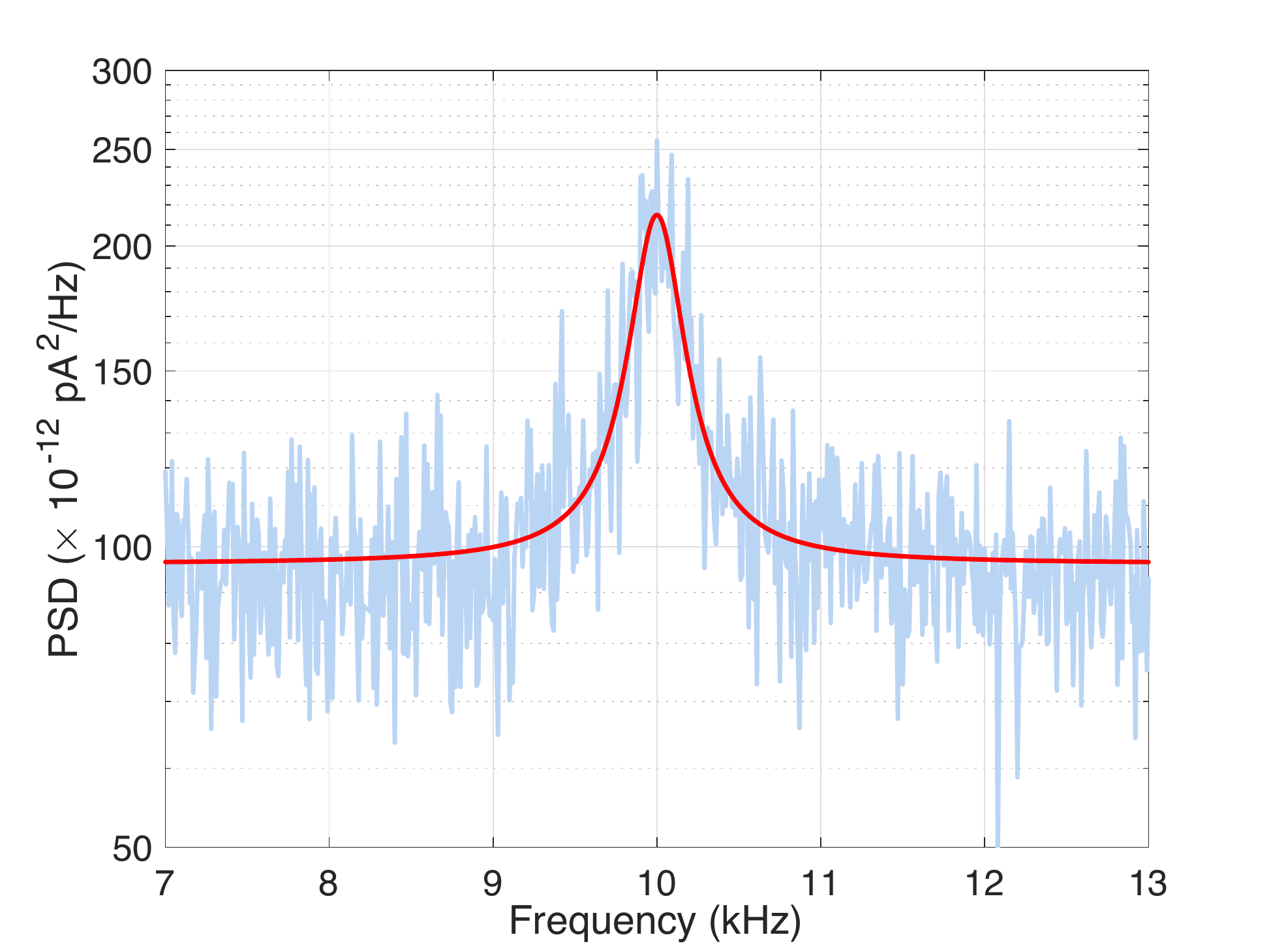}
\captionof{figure}{%
Noise spectroscopy of the meter signal \cite{Lucivero2016a,Lucivero2016b} used to characterize  parameters $\{T_2, Q_{y},Q_{z},R  \}$.  The averaging time of the shown spectrum corresponds to $\unit{50}{\second}$ .}
\label{fig:sns-psd}
\end{minipage}
\hfill
\begin{minipage}[c]{0.45\linewidth}
\centering
\includegraphics[width=\columnwidth]{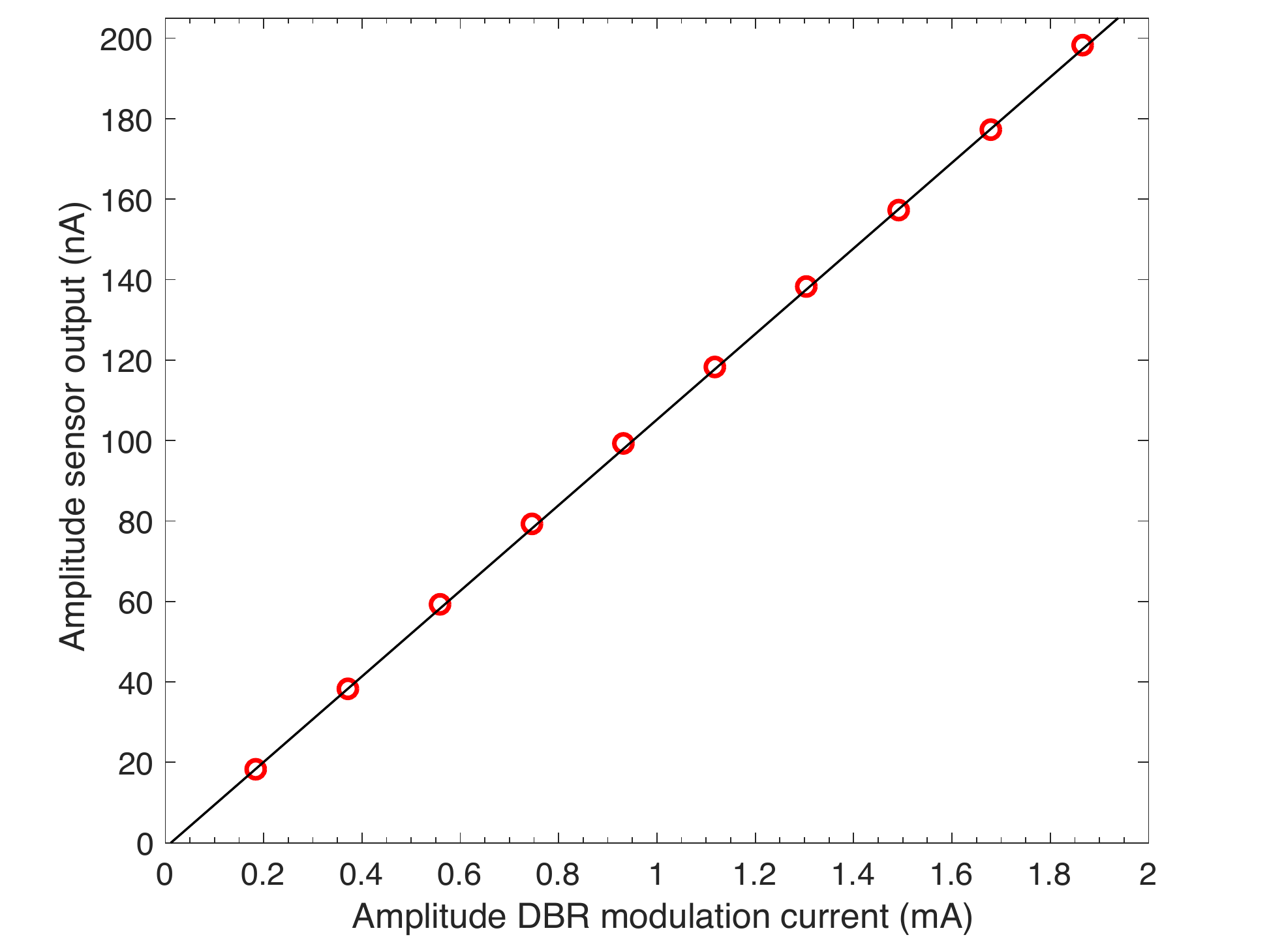}
\caption{%
Linear response of atomic sensor output to a sinusoidal drive. The data points, shown in red, correspond to the observed sensor output obtained at a probe optical power $P = \unit{100}{\micro\watt}$. The solid line (in black) corresponds to a linear fit of the data. From the slope $b = \unit{106.4 \pm 0.4}{\nano\ampere/\milli\ampere} $ of the linear fit we extract the effective coupling constant $\gP$ via $b  =\gD \gP$.}
\label{fig:pump_characterization}
\end{minipage}
\end{figure}

\subsection{Sensor characterization}

We use noise spectroscopy of the meter signal (c.f.~\cite{Lucivero2016a,Lucivero2016b}) to determine the sensor parameters $\{T_2, Q_{y},Q_{z}, R\}$. \figref{fig:sns-psd} shows a typical spectrum of the sensor output at the operating conditions of our experiment, yet in the situation when the pump beam is not coupled to the atomic ensemble. 

The dynamical model of the atomic sensor---the spin dynamics and detection process 
described by \eqnsref{eq:dyns_atoms}{eq:I_meas} of the main text, respectively---predicts
the power spectrum, $S(\omega)$, to follow:
\begin{equation}
	S(\omega) =  S_\t{ph} + \frac{S_\t{at}}{(1/T_2)^2 + (\omega -\omega_0)^2}, 
	\label{Eq:sns-model}
\end{equation} 
which we fit to the observed spectrum with the free parameters of the model being $\{S_\t{ph},S_\t{at},T_2,\omega_0\}$ (see the red curve in \figref{fig:sns-psd}), with $\omega_0 = \omega_L$. From the fit we directly obtain the spin coherence time $T_2$ and straightforwardly determine the variance of the stochastic increments of the spin-noise vector $\d\vB{w}_t^{(\t J)}$ (see \eqnref{eq:dyns_atoms}), $\gD^2Q_{y} = \gD^2Q_{z} = 2S_\t{at}/T_2$, as well as the variance $R^\Delta = R/\Delta$ (with $R = S_\t{ph}$) of the photon shot-noise (see Eq. \eref{eq:I_meas}), following the methods described in \refcite{Lucivero2016a}. \tabref{table:parameters} summarizes the values of the fitted model parameters. 

\begin{table}[!h]
\begin{tabular}{r r r}
\hline\hline 
Parameter & Value  & Unit \\ [0.1ex] 
\hline 
$ (2\pi T_2)^{-1}$ & $182 \pm 4.1 $    & \unit{}{\hertz} \\
$\omega_0/(2\pi)$ & $9999.8 \pm 2.9 $ & $\unit{}{\hertz}$ \\
$S_\t{at}$ & $ 118.7 \pm   1.9  $ & $  (\rm p A)^{2}/\unit{}{Hz}$  \\ 
$S_\t{ph}$ & $ 96.0 \pm   0.3 $ & $   (\rm p A)^{2}/\unit{}{Hz}$  \\ 
\hline 
\end{tabular}
\caption{Dynamical model parameters estimated from 
the spin-noise spectroscopy signal.}
\label{table:parameters}
\end{table}

To calibrate the atomic response to the pump light beam we couple the pumping light to the ensemble and record the sensor output as a function of the amplitude of a resonant sinusoidal drive ($\omega_P = \omega_L = \unit{10}{\kilo\hertz}$) applied to the injection current of the DBR pump-light laser. In \figref{fig:pump_characterization} we plot the observed amplitude (red data points) of the sensor output as a function of the DBR current modulation, as well as a linear fit to the data (solid line). From the slope $b = \unit{106.4 \pm 0.4}{\nano\ampere/\milli\ampere} $ of the linear fit we extract the effective coupling constant $\gP$.
%

\section{Kalman Filter as waveform estimator}
\label{app:KF}

In this appendix, we describe the construction of the \emph{Kalman Filter} (KF)as the \emph{optimal waveform estimator} for system 
and observation (measurement)  linear-Gaussian dynamical models. We start by considering the \emph{continuous-continuous} model, in which both the system and
observations dynamics are described by continuous-time processes, and present the \emph{Kalman-Bucy Filter} (KBF) that is then guaranteed to yield
waveform estimates that minimise the \emph{mean squared error} (MSE) at any time. We then consider the case of time-discrete observations,
i.e., the \emph{continuous-discrete} model, for which the optimal estimator is provided by the \emph{hybrid Kalman Filter} (HKF) that
we utilise in our experiment, as described in the main text. An interested reader is referred for more details to textbooks on classical
filtering theory, e.g., by \citet{Jazwinski1970} or \citet{vanTrees2013}.

\subsection{Continuous-continuous model and the Kalman-Bucy Filter}
\label{sec:cc_model}
%

\subsubsection{Continuous system and measurement dynamics}
Let us consider the case of system dynamics being described by a stochastic process with Gaussian noise, which formally 
corresponds to a time-varying Langevin equation (see, e.g., \cite{gardiner1985}) that dictates the evolution of the 
system \emph{state vector}, $\vB{x}_t$, i.e., \cite{vanTrees2013}:
\BE
\frac{\d\vB{x}_t }{\d t}=\mB{F}_t \vB{x}_t +\mB{\Gamma}_t \vB{u}_t +\mB{G}_t \vB{w}_t ,
\label{eq:state_dyn_lang}
\EE
where $\mB{F}_t$, $\mB{\Gamma}_t$, $\mB{G}_t$ are generally time-dependent matrices, while 
$\vB{u}_t$ is a deterministically evolving vector, e.g., representing external force applied to the system. The initial conditions 
are fixed by specifying the mean state vector and its coviarance matrix at the initial time $t_{0}$, i.e., $\boldsymbol{\mu}_0:=\mean{\vB{x}_{t_{0}}}$ and 
$\cov_{\vB{x}_{t_0}}:=\mean{\vB{x}_{t_{0}}\vB{x}_{t_{0}}^{T}}$, respectively, what then determines the initial Gaussian probability distribution
of the state vector as $\vB{x}_0\sim\cN(\boldsymbol{\mu}_0,\cov_{\vB{x}_{t_0}})$. On the other hand, the measurement outcomes are described by 
the \emph{observations vector}, $\vB{z}_t$, which is assumed to be linearly related at all times to the state vector and to experience an independent stochastic Gaussian noise, i.e.:
\BE
\vB{z}_t =\mB{H}_t \vB{x}_t +\vB{v}_t
\label{eq:obs_dyn_lang}
\EE
with the matrix $\mB{H}_t$ being again in principle time-dependent. In \eqnsref{eq:state_dyn_lang}{eq:obs_dyn_lang}, $\vB{w}_t $ and $\vB{v}_t $ denote the noise vectors---vectors with components consisting of 
stochastic (Wiener) \emph{white-noise terms} \cite{gardiner1985}---such that for all $t$ and $s$:
\AL{
\mean{\vB{w}_t}=\mean{\vB{v}_t}=0, & \qquad\mean{\vB{w}_t \vB{v}_{s}^{T}}=\mean{\vB{v}_t \vB{w}_{s}^{T}}=0, \label{eq:white-noise_terms_1}\\
\mean{\vB{w}_t \vB{w}_{s}^{T}}=\mB{Q}_t \,\delta(t-s), & \qquad\mean{\vB{v}_t \vB{v}_{s}^{T}}=\mB{R}_t \,\delta(t-s), \label{eq:white-noise_terms_2}
}%
where $\mB{Q}_t $ and $\mB{R}_t $ are the noise (symmetric) covariance matrices that fully determine the properties of corresponding
Gaussian fluctuations, i.e., $\vB{w}_t \sim\cN\!\left(0,\mB{Q}_t \right)$ and $\vB{v}_t \sim\cN\!\left(0,\mB{R}_t \right)$,
and have a diagonal form, $\mB{Q}_t=\t{diag}\!\left\{\{Q_i(t)\}_i\right\}$ and $\mB{R}_t=\t{diag}\!\left\{\{R_i(t)\}_i\right\}$, assuming the distinct components of the noise 
vectors to be uncorrelated. 

As the white-noise terms are ill-defined in the $\d t\to0$ limit, in order to formally rewrite \eqnsref{eq:state_dyn_lang}{eq:obs_dyn_lang} 
as stochastic differential equations, one must employ the It\={o} (or Stratonovich---not considered here) calculus, within which they read, respectively 
\cite{gardiner1985}:
\AL{
\d\vB{x}_t  & = \mB{F}_t \vB{x}_t {\d}t+\mB{\Gamma}_t \vB{u}_t {\d}t+\mB{G}_t \d\vB{w}_t, \label{eq:state_dyn_Ito}\\
\d\vB{z}_t  & =\mB{H}_t \vB{x}_t {\d}t+\d\vB{v}_t, \label{eq:obs_dyn_Ito}
}%
where now $\d\vB{w}_t=\{\sqrt{Q_i(t)} \d W_i(t)\}_i$ and $\d\vB{v}_t=\{\sqrt{R_i(t)} \d W_i(t)\}_i$ constitute vectors of \emph{Wiener increments}, ${\d}W_{i}(t)$, which by the It\={o} rules must satisfy $\d W_{i}(t) \d W_{j}(t)=\delta_{ij}{\d}t$ and ${\d}W_{i}(t)^{2+k}={\d}W_{i}(t){\d}t=0$ for all $k>0$. Moreover, \eqnsref{eq:white-noise_terms_1}{eq:white-noise_terms_2}
specifying the noise properties can then be rewritten in terms of the It\={o} differentials as:
\BEA
\mean{\d\vB{w}_t }=\mean{\d\vB{v}_t}=0,
&& \qquad 
\mean{\d\vB{w}_t,\d\vB{v}_{s}^{T}}=\mean{\d\vB{v}_t \d\vB{w}_{s}^{T}}=0, 
\label{eq:Wiener_terms_1}\\
\mean{\d\vB{w}_t \d\vB{w}_{s}^{T}}=\mB{Q}_t \,\delta(t-s)\,\d t,
&&\qquad
\mean{\d\vB{v}_t\d\vB{v}_{s}^{T}}=\mB{R}_t \,\delta(t-s)\,\d t.
\label{eq:Wiener_terms_2}
 \EEA

\subsubsection{Estimator minimising the MSE given the observation record:~the KBF}
For given process \eref{eq:state_dyn_Ito} and observation \eref{eq:obs_dyn_Ito}
models, we would like to construct the most accurate estimate of the state vector at time $t$, i.e., $\vB{x}_t$, basing on the measurement record of 
all observations collected in the past, i.e., $\left\{ \vB{z}_{\tau}\right\} _{\tau<t}$.
Such an \emph{estimator} may be formally defined as a random variable $\est{\vB{x}}_t :=f_t \!\left(\left\{ \vB{z}_{\tau}\right\} _{\tau<t}\right)$
determined by some function $f_t$ that is designed to most efficiently interpret the observation record and predict $\vB{x}_t$ given particular dynamical 
models \eref{eq:state_dyn_Ito} and \eref{eq:obs_dyn_Ito}. Let us define for a given estimator the \emph{error covariance matrix} that quantifies its 
deviation from the true state vector $\vB{x}_t$ at time $t$ as 
\BE
\cov_t :=\mean{(\vB{x}_t -\est{\vB{x}}_t )(\vB{x}_t -\est{\vB{x}}_t )^{T}}.
\label{eq:cov_mat}
\EE
One seeks the optimal estimator minimising some \emph{figure of merit} that quantifies the precision, i.e., the average distance of the estimator from the true state vector: 
\BE
\tr{\mB{W}\,\cov_t} =\mean{(\vB{x}_t -\est{\vB{x}}_t )^{T}\mB{W}(\vB{x}_t -\est{\vB{x}}_t )},
\label{eq:est_cost}
\EE
where $\mB{W}$ is a weight matrix specifying contributions of each vector element to the overall estimation error. Choosing
$\mB{W}=\openone$, in which case all vector components contribute equally, \eqnref{eq:est_cost} simplifies to the \emph{mean squared error} (MSE):
\BE
\t{MSE}(t):=\tr{\cov_t} =\mean{(\vB{x}_t -\est{\vB{x}}_t )^{T}(\vB{x}_t -\est{\vB{x}}_t )}=\mean{\left|\vB{x}_t -\est{\vB{x}}_t \right|^{2}}.
\label{eq:MSE}
\EE
Then, (see, e.g., \cite{Jazwinski1970,vanTrees2013}) by explicitly differentiating \eqnref{eq:MSE} with respect to $\est{\vB{x}}_t $, 
one may prove that the optimal estimator minimising the MSE is the \emph{mean of the posterior distribution} 
$p(\vB{x}_{t+\delta t}|\left\{\vB{z}_{\tau}\right\}_{\tau<t})$,
which describes the probability of system being in the state $\vB{x}_{t+\delta t}$ at time $t+\delta t$ given the past observation record $\left\{ \vB{z}_{\tau}\right\} _{\tau<t}$.
Hence, the optimal estimator at time $t+\delta t$ may always be formally written as 
\BE
\est{\vB{x}}_{t+{\delta}t}=\int_t ^{t+\delta t}\!\!\!D\vB{x}\;\vB{x}_{t+\delta t}\;p(\vB{x}_{t+\delta t}|\left\{ \vB{z}_{\tau}\right\} _{\tau<t}),
\label{eq:posterior_mean}
\EE
where $\int_t ^{t+\delta t}D\vB{x}$ denotes averaging over the fluctuations of the state 
vector occurring within the most recent interval, $[t,t+\delta t]$, after recording the last observation. 

However, in the case of linear-Gaussian process and observation models---in particular, \eqnsref{eq:state_dyn_Ito}{eq:obs_dyn_Ito}---such an optimal estimator 
can be shown to satisfy an ordinary differential equation, i.e., the \emph{Kalman-Bucy equation} \cite{Kalman1960, Kalman1961}:
\BE
\frac{\d\est{\vB{x}}_t }{{\d}t}=\mB{F}_t \est{\vB{x}}_t +\mB{\Gamma}_t \vB{u}_t +\mB{K}_t \left(\vB{z}_t -\mB{H}_t \est{\vB{x}}_t \right),
\label{eq:KB_eq}
\EE
where the term in brackets is known as the \emph{innovation}, i.e.,  
\BE
\innov_t :=\vB{z}_t - \est{\vB{z}}_t 
\quad\text{with}\quad
\est{\vB{z}}_t:=\mB{H}_t \est{\vB{x}}_t 
\label{eq:innovation}
\EE
representing then the effective estimate of the observation at time $t$, also provided by the estimator construction.
The matrix $\mB{K}_t $ in \eqnref{eq:KB_eq} is the so-called \emph{Kalman gain}:
\BE
\mB{K}_t:=\cov_t \mB{H}_t ^{T}\mB{R}_t ^{-1},
\label{eq:Kalman_gain}
\EE
which formally depends on the error covariance matrix $\cov_t $ of the corresponding optimal estimator. 
Nevertheless, $\mB{K}_t$ may be determined independently of $\est{\vB{x}}_t$, as $\cov_t $ can be shown to optimally 
fulfil the \emph{variance equation} \cite{Kalman1960, Kalman1961}:
\BE
\frac{\d\cov_t }{{\d}t}=\mB{F}_t \cov_t +\cov_t \mB{F}_t ^{T}+\mB{G}_t \mB{Q}_t \mB{G}_t ^{T}-\cov_t \mB{H}_t ^{T}\mB{R}_t ^{-1}\mB{H}_t \cov_t,
\label{eq:var_eq}
\EE
which constitutes an example of \emph{matrix Riccatti} (ordinary differential) \emph{equation} that,
despite being non-linear in $\cov_t$, can always be efficiently solved, at least numerically \cite{Jazwinski1970}.
Combined solutions to \eqnsref{eq:KB_eq}{eq:var_eq} provide the optimal $\est{\vB{x}}_t $ as an integral of \eqnref{eq:KB_eq} over the 
observations $\vB{z}_t $ collected in the past. Such an estimator (which, however, often can only be computed numerically) is termed 
as the \emph{Kalman-Bucy filter} (KBF) \cite{vanTrees2013}.

\subsubsection{Steady-state solution of KBF}
\label{sec:sss_kbf}

Under quite general conditions (see \refcite{Kalman1961}) and, in particular, when dealing with time-invariant 
dynamical models (when the evolution models \eref{eq:state_dyn_Ito} and \eref{eq:obs_dyn_Ito} are described 
by time-invariant $\mB{F}$, $\mB{G}$, $\mB{H}$, $\mB{R}$ and $\mB{Q}$),
the solution of \eqnref{eq:var_eq} must \emph{stabilize} with time, so that $\d\cov_t /\d t\to\vB{0}$ as $t\to\infty$.
In such an asymptotic regime, the error covariance matrix approaches a constant matrix, i.e., the \emph{steady-state solution} $\cov_\infty$, 
for which the r.h.s.~of \eqnref{eq:var_eq} vanishes. Hence,  $\cov_\infty$ corresponds to the solution 
of the \emph{continuous algebraic Riccatti equation} (CARE) \cite{vanTrees2013},
\BE
\vB{0}=\mB{F} \cov_\infty +\cov_\infty \mB{F} ^{T}+\mB{G} \mB{Q} \mB{G} ^{T}-\cov_\infty \mB{H} ^{T}\mB{R} ^{-1}\mB{H} \cov_\infty,
\label{eq:ARE_cc}
\EE
which then also determines the asymptotic value attained by the filter gain:~$\mB{K}_\infty=\cov_\infty \mB{H}^{T}\mB{R}^{-1}$.
As \eqnref{eq:ARE_cc} constitutes a matrix equation that is quadratic in $\cov_\infty$, it is typically hard to find its analytical 
solution. However, efficient numerical methods are well-established, e.g., by employing the Schur decomposition method \cite{Laub1979}. 
Crucially, the steady-state solution, $\cov_\infty$, quantifies the overall performance
of the KBF---the minimal MSE \eref{eq:MSE}, $\lim_{t\to\infty}\t{MSE}(t)=\tr{\cov_\infty}$, that may be 
attained for a particular continuous linear-Gaussian model (\ref{eq:state_dyn_Ito}-\ref{eq:obs_dyn_Ito})
over large time-scales, i.e., when the waveform estimation procedure stabilises reaching its fundamental limits. 

However, as in our atomic sensor experiment the measurements are taken at non-negligible time intervals---the sampling period $\Delta$ introduced in \appref{sec:det_photocurr}---in what follows we must generalise the above derivation accounting explicitly for the time-discrete character of the observation model \eref{eq:obs_dyn_Ito}---see \eqnref{eq:I_meas} of the main text. Nevertheless, let us emphasize that the solutions obtained for such a time-discrete observation model must converge to the ones provided by the KBF and the CARE \eref{eq:ARE_cc} in the limit of sufficiently frequent measurements, i.e., $\Delta \to 0$.

\subsection{Continuous-discrete model and the Hybrid Kalman Filter}
\label{sec:cd_model}

\subsubsection{Continuous system but discrete measurement dynamics}
When the measurements are performed in finite time-steps dictated by the sampling interval (period) $\Delta$, the observations must be formally described 
by a sequence of outcomes:~$\left\{ \vB{z}_{k}\right\} _{k=0}^{\left\lfloor t/\Delta\right\rfloor }$ with $t_{k}=k\Delta+t_{0}$ and $k\in\mathbb{N}$. 
Note that, for simplicity, we employ a notation in which any time-discrete quantity evaluated at time $t_{k}$ is labelled by the subscript $k$, e.g.,
$\vB{z}_{k}\equiv\vB{z}_{t_{k}}$. In such a time-discrete observation setting, the dynamics are described by a ``hybrid'' \emph{continuous-discrete} model, 
in which the system evolves according to a time-continues process \eref{eq:state_dyn_Ito} while observations must be described employing the Langevin 
formulation \eref{eq:obs_dyn_lang}:
\AL{
\d\vB{x}_t  & =\mB{F}_t \vB{x}_t {\d}t+\mB{\Gamma}_t \vB{u}_t {\d}t+\mB{G}_t \d\vB{w}_t ,\label{eq:cd_model_state}\\
\vB{z}_{k} & =\mB{H}_{k}\vB{x}_{k}+\vB{v}_{k},\label{eq:cd_model_obs}
}%
with the stochastic vector $\vB{v}_{k}\sim\cN(0,\mB{R}_k^\Delta)$ representing now, in contrast to time-continuous \eqnref{eq:obs_dyn_lang}
with $\vB{v}_{t}\sim\cN(0,\mB{R}_t)$, 
a $k$-sequence defined by a \emph{discrete} white-noise random process \cite{gardiner1985}.
Crucially, the Langevin term describing the observation noise fluctuates now with covariance 
$\mB{R}_k^\Delta:=\mB{R}_{t_k}/\Delta$, so that  the observation model \eref{eq:cd_model_obs}consistently converges to \eqnref{eq:obs_dyn_lang} in the continuous measurement limit of $\Delta\to0$ \cite{bar2004} (what then directly follows from \eqnref{eq:discretised_obs_noise}).

\subsubsection{Estimator minimising the MSE given the observation record:~the HKF}
\label{sec:HKF_constr}
As discussed in the previous section, for any inference model the mean of the posterior distribution \emph{always} constitutes the optimal 
estimator minimising the MSE. Hence, we may now formally define the optimal estimator by simply rewriting \eqnref{eq:posterior_mean} and accounting 
for the time-discrete character of the observations:
\BE
\est{\vB{x}}_{k|k-1}=\int_{k-1}^{k}\!\!D\vB{x}\;\vB{x}_{k}\;p(\vB{x}_{k}|\left\{ \vB{z}_{k'}\right\} _{k'=0}^{k-1}),
\label{eq:puigdemont}
\EE
where $\int_{k-1}^{k}D\vB{x}\equiv\int_{t_{k-1}}^{t_{k}}D\vB{x}$ denotes now the averaging over the state fluctuations occurring during 
the $[t_{k-1},t_k]$ interval just before the $k$th observation is recorded. In the standard notation adopted above \cite{vanTrees2013}, $\est{\vB{x}}_{k|k-1}$represents
the optimal estimator of the state vector at time $t_{k}$ given the past observation record $\left\{ \vB{z}_{k'}\right\} _{k'=0}^{k-1}$, while 
$\est{\vB{x}}_{k|k}$ denotes the estimator at time $t_{k}$ that, however, has already been \emph{updated} basing on the observation 
$\vB{z}_{k}$. Similar notation is used for the error covariance matrix \eref{eq:cov_mat} of $\est{\vB{x}}_{k|k-1}$ and $\est{\vB{x}}_{k|k}$,
corresponding then to $\cov_{k|k-1}$ and $\cov_{k|k}$, respectively.

As the continuous-discrete model is described by \eqnsref{eq:cd_model_state}{eq:cd_model_obs}that are still linear-Gaussian processes, the corresponding optimal estimator minimising the MSE is constructed in an analogous fashion to \eqnref{eq:posterior_mean} defining the KBF, but in an explicit two-step \emph{prediction} and \emph{update} procedure \cite{Jazwinski1970,vanTrees2013} due to $\delta t\to0$ limit being no longer valid in \eqnref{eq:puigdemont}. 
Such a construction is then optimal, as due to lack of any outcome information in between the measurements the estimator within such time-intervals can only be evolved according to the system dynamics. At times $t_{k}=t_{0}+k\Delta$, on the other hand, 
it must be just updated basing on a particular outcome registered. Consequently, the optimal estimator is then called the 
\emph{hybrid Kalman filter} (HKF) and it consistently converges to the KBF---the solution of \eref{eq:KB_eq}---in the $\Delta\to0$ limit, 
in which the time-discrete Langevin equation \eref{eq:cd_model_obs} converges to its time-continuous form \eref{eq:obs_dyn_lang}
\cite{bar2004}.

\textbf{Filter initialisation.}
Firstly, however, one must initialise the HKF at time $t_0$ after deciding on an appropriate initial Gaussian distribution, $p(\vB{x}_{0})\sim\cN(\boldsymbol{\mu}_0,\cov_{\vB{x}_0})$, that adequately represents the knowledge about the state vector $\vB{x}_{0}\equiv\vB{x}(t_{0})$ prior to the estimation procedure. 
This corresponds to setting the initial HKF estimates of $\vB{x}_{0|0}$ and the covariance matrix $\cov_{0|0}$ to, respectively:
\BE
\est{\vB{x}}_{0|0}=\int\mathrm{d}\vB{x}_{0}\,p(\vB{x}_{0})\,\vB{x}_{0}=: \boldsymbol{\mu}_0,
\qquad
\cov_{0|0}=\int\!\!\d\vB{x}_{0}\,p(\vB{x}_{0})\,\left(\vB{x}_{0}-\boldsymbol{\mu}_0\right)\left(\vB{x}_{0}-\boldsymbol{\mu}_0\right)^{T}=: \cov_{\vB{x}_{0}}.
\EE
Here, we choose the Gaussian prior to be the distribution optimally inferred from a single observation $\vB{z}_{0}$ taken at the initial time $t_0$ \cite{bar2004}. In particular, we set the mean to $\boldsymbol{\mu}_0 = \mB{H}_0^{+}\vB{z}_0$, where $\mB{H}_0^{+}$ denotes the pseudoinverse of $\mB{H}_0$ in \eqnref{eq:cd_model_obs} at $t_0$, while the variance to $\cov_{\vB{x}_0} =\mB{Q}_0 +\mB{H}_0^{+}\mB{R}_0^\Delta(\mB{H}_0^{+})^{\rm T }$
in order to account for the uncertainty in filter initialisation due to intrinsic (unconditional) system and detection noises 
(determined by $\mB{Q}_0$ and $\mB{R}_0^\Delta$  of \eqnsref{eq:cd_model_state}{eq:cd_model_obs}, respectively).

\textbf{Prediction step \emph{($\est{\vB{x}}_{k-1|k-1}\to\est{\vB{x}}_{k|k-1}$ and $\cov_{k-1|k-1}\to\cov_{k|k-1}$).}}
In order to perform the prediction step, let us define the \emph{transition
matrix}, $\mB{\Phi}$, as the solution of the non-stochastic part of the state-vector continuous dynamics 
\eref{eq:cd_model_state} with also the deterministic term $\vB{u}_t $ being disregarded.
In particular, we define $\mB{\Phi}_{t,s}$ (for a general time interval $[s,t]$) to be the matrix solution of
\BE
\frac{\d\mB{\Phi}_{t,s}}{{\d}t}=\mB{F}_t \mB{\Phi}_{t,s}
\EE
that must also satisfy $\mB{\Phi}_{\tau,\tau}=\openone$ for all $\tau\ge0$. Hence, the transition matrix may be formally written as
\BE
\mB{\Phi}_{t,s}=\mathcal{T}_{\leftarrow}\exp\!\left[\int_{s}^{t}\d\tau\mB{F}_{\tau}\right]=\sum_{n=0}^{\infty}\frac{1}{n!}\,\mathbf{S}_{s,t}^{(n)}
\label{eq:state_trans_matrix}
\EE
where 
\BE
\mathbf{S}_{s,t}^{(n)}:=\mathcal{T}_{\leftarrow}\int_{s}^{t}\!\!\int_{s}^{t_{1}}\!\!\!\dots\!\int_{s}^{t_{n-1}}\!\!\!\!\mB{F}_{t_{1}}\mB{F}_{t_{2}}\dots\mB{F}_{t_{n}}\,{\d}t_{1}{\d}t_{2}\dots{\d}t_{n}=\int_{s}^{t}\!\mB{F}_{t_{1}}\int_{s}^{t_{1}}\!\!\mB{F}_{t_{2}}\dots\int_{s}^{t_{n-2}}\!\!\!\!\mB{F}_{t_{n-1}}\int_{s}^{t_{n-1}}\!\!\!\!\mB{F}_{t_{n}}\,{\d}t_{n}{\d}t_{n-1}\dots{\d}t_{2}{\d}t_{1}.
\label{eq:madafaka}
\EE
$\mathcal{T}_{\leftarrow}$ in \eqnref{eq:state_trans_matrix} denotes the time-ordering operation, as defined in 
\eqnref{eq:madafaka}, but may always be ignored in case the $\mB{F}$-matrices commute at different time-instances, i.e.,
when $\left[\mB{F}_t ,\mB{F}_{s}\right]=0$ for all $t$ and $s$. 
Moreover, in the case when the $\mB{F}$-matrix is time-independent,
so that $\mB{\Phi}_{t,s}\equiv\mB{\Phi}_{t-s}=\e^{\mB{F}(t-s)}$,
the transition matrix for any $\Delta$-interval  $[t_{k-1},t_{k}]$ is the same and reads 
$\mB{\Phi}:=\mB{\Phi}_\Delta=\e^{\mB{F}\Delta}$.

With help of $\mB{\Phi}_{t,s}$, we can construct $\est{\vB{x}}_{k|k-1}$ 
as a function of $\est{\vB{x}}_{k-1|k-1}$ by integrating the estimator over the interval $[t_{k-1},t_{k}]$ according to the 
deterministic part of the system dynamics \eref{eq:cd_model_state},
\BE
\est{\vB{x}}_{k|k-1}=\mB{\Phi}_{t_{k},t_{k-1}}\est{\vB{x}}_{k-1|k-1}+\int_{t_{k-1}}^{t_{k}}\mB{\Phi}_{t_{k},\tau}\mB{\Gamma}_{\tau}\vB{u}_{\tau}\d\tau,
\label{eq:HKF_predict}
\EE
while also adequately propagating the estimator covariance matrix:
\BE
\cov_{k|k-1}=\mB{\Phi}_{t_{k},t_{k-1}}\cov_{k-1|k-1}\mB{\Phi}_{t_{k},t_{k-1}}^{T}+\mB{Q}_k^\Delta,
\label{eq:cov_mat_predict}
\EE
where
\BE
\mB{Q}_k^\Delta:=\int_{t_{k-1}}^{t_{k}}\mB{\Phi}_{t_{k},\tau}\mB{G}_{\tau}\mB{Q}_{\tau}\mB{G}_{\tau}^{T}\mB{\Phi}_{t_{k},\tau}^{T}\d\tau,
\label{eq:Q_Delta}
\EE
now represents the \emph{effective} covariance matrix of the system noise, which importantly accounts for 
the finite sampling period, $\Delta$,
of the time-discrete observation model \eref{eq:cd_model_obs} (see also \eqnref{eq:cov_pred} of the main text). 

Note that the expression \eref{eq:HKF_predict} for the predicted HKF constitutes the integral solution 
to the Kalman-Bucy equation \eref{eq:KB_eq} for the $[t_{k-1},t_k]$ interval with the observation-based 
updating completely ignored, i.e., the Kalman gain set to zero ($\mB{K}_{t}=\mzero$) in \eqnref{eq:KB_eq}. Similarly, the error covariance matrix  
\eref{eq:cov_mat_predict} of the prediction satisfies the variance equation \eref{eq:var_eq} with the last term ignored, which in the case of 
the continuous-continuous model  stood for the observation-based correction to the estimator.

\textbf{Update step \emph{($\est{\vB{x}}_{k|k-1}\to\est{\vB{x}}_{k|k}$  and $\cov_{k|k-1}\to\cov_{k|k}$)}.}
In order to incorporate into the estimator the $k$th outcome, $\vB{z}_k$, one simply adds to the prediction 
the $\vB{z}_k$-based innovation multiplied by the Kalman gain, i.e.,
\BE
\est{\vB{x}}_{k|k}=\est{\vB{x}}_{k|k-1}+\mB{K}_{k} \innov_{k},
\label{eq:est_update}
\EE
where the innovation and the Kalman gain now read, respectively:
\BE
\innov_{k}=\vB{z}_{k}-\est{\vB{z}}_k
\qquad\text{and}\qquad
\mB{K}_{k}=\cov_{k|k-1}\mB{H}_{k}^{T}\mB{S}_{k}^{-1}.
\label{eq:innov_gain_update}
\EE
As before, $\est{\vB z}_k:=\mB{H}_{k}\est{\vB{x}}_{k|k-1}$ should be interpreted above 
as the filter-based prediction of the $k$th outcome value. For convenience, we have now explicitly defined 
the covariance matrix for the $k$th innovation $\innov_k$ above as
\BE
	\vecb{S}{k} : = \mean{\innov_k \innov_k^T} = \mB{R}_k^\Delta + \mB{H}_{k}\cov_{k|k-1}\mB{H}_{k}^T,
\EE
whose behaviour, when explicitly computed and analysed for particular data, can also be utilised to verify the validity and accuracy 
of processes \eref{eq:cd_model_state} and \eref{eq:cd_model_obs} describing the system and observation real dynamics \cite{bar2004} (see, in particular, \figref{fig:OU-signal}(e) of the main text for the case the atomic sensor under study).

Finally, it is then straightforward to show that the estimator transformation \eref{eq:est_update} results in the following update 
of its error covariance matrix \eref{eq:cov_mat}: 
\BE
\cov_{k|k}=\left(\openone-\mB{K}_{k}\mB{H}_{k}\right)\cov_{k|k-1}
\label{eq:cov_mat_update}
\EE
with the Kalman gain defined in \eqnref{eq:innov_gain_update}.

\subsubsection{Steady-state solution for the HKF}
Similarly to the case of the KBF, when considering the continuous-discrete models but with time-invariant  $\mB{F}$, $\mB{G}$, $\mB{H}$, $\mB{R}^\Delta$ 
and $\mB{Q}^\Delta$ in \eqnsref{eq:cd_model_state}{eq:cd_model_obs} (for which then also $\mB{\Phi}=\e^{\mB{F}\Delta}$), the HKF approaches a 
\emph{steady-state solution} as $k\to\infty$ \cite{vanTrees2013}. However, as the discrepancy between predicted and updated state values 
can be shown to be persistent also in the steady-state regime, one must define separately the constant values approached by the corresponding
error covariance matrices for the prediction and update steps, respectively, as follows:
\BE
\covSSp := \lim_{k\to\infty} \cov_{k|k-1}
\qquad\text{and}\qquad
\covSSu :=  \lim_{k\to\infty} \cov_{k|k}, 
\EE

The form of $\covSSp$ can be determined by substituting into \eqnref{eq:cov_mat_predict} the expression for $\cov_{k-1|k-1}$ according to \eqnref{eq:cov_mat_update}, 
which then in the $k\to\infty$ limit (in which $\cov_{k|k-1}\approx\cov_{k-1|k-2}\,\to\,\covSSp$) yields the \emph{discrete algebraic Riccatti equation} (DARE), i.e., 
the equivalent of \eqnref{eq:ARE_cc} for the continuous-discrete case \cite{vanTrees2013}:
\BE
\covSSp=\mB{\Phi}\covSSp\mB{\Phi}^T - \mB{\Phi}\covSSp \mB{H}^T\left(\mB{R}^\Delta+\mB{H}\covSSp\mB{H}^T\right)^{-1}\mB{H}\covSSp\mB{\Phi}^{T} + \mB{Q}^\Delta,
\label{eq:ARE_cd_p}
\EE
The asymptotically attained value $\covSSp$ consequently allows us to compute the Kalman gain and the innovation covariance matrix for 
the steady-state regime, i.e.:
\BE
\mB{K}_{\t{ss}}=\covSSp\mB{H}^{T}\mB{S}_{\t{ss}}^{-1}
\quad\text{and}\quad
\mB{S}_{\t{ss}}=\mB{R}^\Delta + \mB{H}\covSSp\mB{H}^T,
\label{eq:gain_ss}
\EE
so that the steady-state error covariance matrix for the update step can then be found using \eqnref{eq:cov_mat_update}:
\BE
\covSSu=\left(\openone-\mB{K}_{\t{ss}}\mB{H}\right)\covSSp.
\label{eq:ARE_cd_u}
\EE

\section{Applying the HKF to the atomic sensor}
\label{sec:KF_atoms}
%

\subsection{Continuous-discrete model describing the atomic sensor}
The atomic sensor under study constitutes an example of the continuous-discrete dynamical model discussed in the previous section.
In particular, the system evolution \eref{eq:cd_model_state} describes the dynamics of both the ensemble spin-components transversal 
to the magnetic field $B_{0}$ (see \figref{fig:setup} of the main text), $\vB{j}_{t}:=\left[J_{y}(t),J_{z}(t)\right]^{T}$, as well as the 
pump-beam quadratures, $\vB{q}_{t}:=\left[q(t),p(t)\right]^{T}$, representing the estimated waveform. 
Hence, in order to track the evolution, we construct the HKF for the state vector:
\BE
\vB{x}_{t}
=
\vB{j}_{t} \oplus \vB{q}_{t}
=
\left[\begin{array}{cc|cc}J_{y}(t) & J_{z}(t) & q(t) & p(t)\end{array}\right]^{T},
\label{eq:state_vector_exp}
\EE
where, for convenience, we explicitly mark above the splitting of all the vectors and matrices into the relevant atomic-spin and quadrature parts. 
The stochastic-noise contribution in \eqref{eq:cd_model_state} is then given by the ${\d}\vB{w}_{t}$ term introduced in the main text below \eqnref{eq:dyns_pump},
which contains all the corresponding Wiener increments of atoms and light, i.e.,
\BE
\d\vB{w}_{t}=
{\d}\vB{w}_{t}^{(\t{J})}\oplus{\d}\vB{w}_{t}^{(\t{q})}=
\left[\begin{array}{cc|cc}\sqrt{Q_y}{\d}W_{y}(t) & \sqrt{Q_z} \d W_{z}(t) &
\sqrt{Q_q} \d W_{q}(t) & \sqrt{Q_p} \d W_{p}(t)\end{array}\right]^{T},
\EE
with noise intensity fully specified by the (diagonal) noise covariance matrix:
\BE
\mB{Q}=\mB{Q}^{(\t{J})}\oplus\mB{Q}^{(\t{q})}=\t{diag}\left\{ Q_{y},Q_{z}\right\} \oplus\t{diag}\left\{ Q_{q},Q_{p}\right\} .
\label{eq:Q_exp}
\EE
Thus, the full dynamics of the state vector\textemdash encompassing both the evolution of the spin-ensemble as well as the stochastically
driven quadratures (see \eqnsref{eq:dyns_atoms}{eq:dyns_pump} of the main text, respectively)\textemdash corresponds to the special choice 
of the $\mB{F}$-matrix in \eqnref{eq:cd_model_state}:
\BE
\mB{F}_{t}
=
\left[\begin{array}{c|c}
\mB{F}^{(\t{J})} & \mB{F}_{t}^{(\t{J\!\cdot\!q})}\\
\hline \boldsymbol{0} & \mB{F}^{(\t{q})}
\end{array}\right]
= 
\left[\begin{array}{cc|cc}
-\frac{1}{T_{2}} & \omega_{\mathrm{L}} & 0 & 0\\
-\omega_{\mathrm{L}} & -\frac{1}{T_{2}} & \gP\cos\!\left(\omega_{\t{P}}t\right) & \gP\sin\!\left(\omega_{\t{P}}t\right)\\
\hline 0 & 0 & -\kappa_{q} & 0\\
0 & 0 & 0 & -\kappa_{p}
\end{array}\right],
\label{eq:F_matrix}
\EE
and, trivial, $\mB{\Gamma}_{t}=0$, $\mB{G}_{t}=\openone$.

On the other hand, the photocurrent detection described in \eqnref{eq:I_meas} of the main text directly translates onto the 
time-discrete observation model \eref{eq:cd_model_obs} with the observation vector $\mathbf{z}_{k}$ being just a scalar 
representing the photocurrent measured at time $t_{k}$. In particular, the atomic-sensor setup corresponds then to just choosing in \eqnref{eq:cd_model_obs}:
\BE
\mathbf{z}_{k}\equiv z_k=I(t_{k}),\quad\vB{v}_{k}\equiv v_k=\xi_\t{D}(t_{k}),
\quad
\mB{H}_k\equiv \mB{H}=\left[\begin{array}{cc|cc}0 & g_{\t{D}} & 0 & 0\end{array}\right],
\label{eq:obs_model_exp}
\EE
and fixing the variance of the noise-term $v_{k}$ to the (scalar) intensity of the detection noise, i.e., $v_k\sim \cN(0,R^\Delta)$,
so that consistently with \eqnref{eq:discretised_obs_noise} for all $k$ and $k'$:~$\mean{v_{k}v_{k'}}=\mean{\xi_\t{D}(t_{k})\xi_\t{D}(t_{k'})}=R^\Delta\,\delta_{kk'}$.

Thanks to the above formulation we may directly apply the construction of the HKF described in the previous sections in order to optimally estimate the state vector \eref{eq:state_vector_exp}. In particular, in accordance with the prescription of \appref{sec:HKF_constr}, we first initialise the HKF at time $t_0$ with an initial (prior) Gaussian distribution $\estRJM{x}{0}{0}\sim\nd(\mB{H}^{-1}\vB{z}_0, \mB{Q} +\mB{H}^{-1}R^\Delta(\mB{H}^{-1})^{\rm T })$ after substituting for the predetermined (see \appref{app:detect_sensor_char}) experimental parameters in all the dynamical matrices. Then, at subsequent time-steps $k>0$, we apply the two-step recursive 
implementation of the HKF to construct the estimator $\vB{x}_{k|k}$ and covariance matrix $\cov_{k|k}$ in an 
efficient manner, while constantly collecting the photocurrent experimental data $z_k$. The dynamical and noise parameters of the atomic sensor
($\omega_{\t{L}}$, $T_{2}$,  $g_{\t{P}}$, $\omega_{\t{P}}$, $\kappa_{p}$, $\kappa_{q}$, $\vB{u}_{t}$, $\gD$, $\mB{Q}$, $R^\Delta$) are either 
predetermined experimentally or pre-set and controlled by us, as discussed in the main text and \appref{app:detect_sensor_char}.

\subsection{Steady-state solution for the HKF applicable to the atomic sensor}
Let us note that in case of the atomic sensor under study all the dynamical matrices ($\mB{G}_{t}$, $\mB{\Gamma}_{t}$,
$\mB{H}_{k}$) in \eqnsref{eq:cd_model_state}{eq:cd_model_obs} are time-independent, with the only exception of $\mB{F}_{t}$,
specified in \eqnref{eq:F_matrix}. However, in the case of the experiment conducted, in which we set the signal quadratures to fluctuate 
according to white-noise of the same intensity, $Q_{q}=Q_{p}=:Q$, also the time-dependence of $\mB{F}_{t}$ can be bridged by 
moving to the ``\emph{rotating frame}'' (RF) of the signal (pump) field which oscillates at the modulation frequency $\omega_{\t{P}}$. 
Defining the corresponding transformation to the RF-picture (denoted by $\bar{\bullet}$) for
any vector $\boldsymbol{\alpha}$ as
\BE
\bar{\boldsymbol{\alpha}}=\mathsf{R}_{\omega_{\t{P}}t}\boldsymbol{\alpha}
\qquad\text{with}\qquad
\mathsf{R}_{\omega_{\t{P}}t}:=\left[\begin{array}{cc}
\cos\omega_{\t{P}}t & \sin\omega_{\t{P}}t\\
-\sin\omega_{\t{P}}t & \cos\omega_{\t{P}}t
\end{array}\right],
\label{eq:RF_trans}
\EE
we can rewrite the stochastic dynamics of the quadratures in the RF (stemming from the \eqnref{eq:dyns_pump} of the main text) as
\BE
\d\bar{\vB{q}}_{t}=\left[\begin{array}{cc}
-\kappa & \omega_{\t{P}}\\
-\omega_{\t{P}} & -\kappa
\end{array}\right]\bar{\vB{q}}_{t}{\d}t+{\d}\bar{\vB{w}}_{t}^{(\t{q})},
\label{eq:quad_dyn_RF}
\EE
where all the vectors are rotated into the RF according to \eqnref{eq:RF_trans}. Note that the covariance matrix of the quadrature-noise 
increments, ${\d}\bar{\vB{w}}_{t}^{(\t{q})}$, remains unchanged, as
\BE
\mean{{\d}\bar{\vB{w}}_{t}^{(\t{q})}{\d}\bar{\vB{w}}_{t}^{(\t{q})T}}=\mathsf{R}_{\omega_{\t{P}}t}\left(\mB{Q}^{(\t{q})}{\d}t\right)\mathsf{R}_{\omega_{\t{P}}t}^{T}
=
\mathsf{R}_{\omega_{\t{P}}t}\left(\t{diag}\{Q,Q\}\right)\mathsf{R}_{\omega_{\t{P}}t}^{T}{\d}t
=
\t{diag}\{Q,Q\}\,{\d}t=\mean{{\d}\vB{w}_{t}^{(\t{q})}{\d}\vB{w}_{t}^{(\t{q})T}}.
\EE

Combining the RF-based quadrature dynamics \eref{eq:quad_dyn_RF} with the unaltered spin-ensemble evolution, we can write the modified dynamics 
of the full state vector \eref{eq:state_vector_exp} as
\BE
{\d}\bar{\vB{x}}_{t}=\mB{F}_{\t{RF}}\bar{\vB{x}}_{t}{\d}t+{\d}\bar{\vB{w}}_{t},
\label{eq:dyn_model_RF}
\EE
where:
\BE
\bar{\vB{x}}_{t}=
\vB{j}_{t}\oplus\bar{\vB{q}}_{t},
\quad\bar{\vB{u}}_{t}=\vB{u}_{t}^{(\t{J})}\oplus\bar{\vB{u}}_{t}^{(\t{q})},
\quad{\d}\bar{\vB{w}}_{t}={\d}\vB{w}_{t}^{(\t{J})}\oplus{\d}\bar{\vB{w}}_{t}^{(\t{q})},
\EE
and the $\mB{F}$-matrix now reads in the RF:
\BE
\mB{F}_{\t{RF}}:=\left[\begin{array}{cccc}
-\frac{1}{T_{2}} & \omega_{\mathrm{L}} & 0 & 0\\
-\omega_{\mathrm{L}} & -\frac{1}{T_{2}} & \gP & 0\\
0 & 0 & -\kappa & \omega_{\t{P}}\\
0 & 0 & -\omega_{\t{P}} & -\kappa
\end{array}\right],
\label{eq:F_RF}
\EE
being, indeed, time-independent. Finally, note that as the observation vector $z_k$, defined in \eqnref{eq:obs_model_exp}
and representing the photocurrent measurement, is coupled only to the atomic-spin dynamics, the observation dynamical model 
remains unaffected by moving to the RF.

Crucially, in the RF picture, we can now write the corresponding DARE \eref{eq:ARE_cd_p} for the atomic sensor, 
in order to determine the steady-state solution of the \emph{prediction}-based error covariance matrix, $\covSSpRF$, i.e.,
\BE
\covSSpRF=\mB{\Phi}\covSSpRF\mB{\Phi}^T - \mB{\Phi}\covSSpRF \mB{H}^T\left(R^\Delta+\mB{H}\covSSpRF\mB{H}^T\right)^{-1}\mB{H}\covSSpRF\mB{\Phi}^{T}+ \mB{Q}^\Delta,
\label{eq:dare_at_sens}
\EE
where the transition matrix reads $\mB{\Phi}_{t}=\e^{\mB{F}_\t{RF}t}$, so that 
according to \eqnref{eq:Q_Delta}:
\BE
\mB{Q}^\Delta=
\mB{\Phi}_\Delta
\left(\int_{0}^{\Delta}\!\!\d\tau \mB{\Phi}_{-\tau}\mB{Q}\mB{\Phi}_{-\tau}^{T}\right)
\mB{\Phi}_\Delta^T,
\EE
while $\mB{Q}$, $\mB{H}$  and $\mB{F}_\t{RF}$ are defined in Eqs.~\eref{eq:Q_exp}, \eref{eq:obs_model_exp} and \eref{eq:F_RF}, respectively.
Consequently, the steady-state solution for the error covariance matrix after the update step, $\covSSuRF$,
is then determined by just substituting the solution of \eqnref{eq:dare_at_sens} into consecutively \eqnsref{eq:gain_ss}{eq:ARE_cd_u}. 

Finally, note that all covariance matrices---in particular, the steady-state solutions---can then be computed in the laboratory frame 
by transforming back from the RF via $\cov=\mathsf{R}_{\t{\omega_\t{P}t}}^{(\t{q,J})}\cov^\t{RF}(\mathsf{R}_{\t{\omega_\t{P}t}}^{(\t{q,J})})^T$,
where $\mathsf{R}_{\t{\omega_\t{P}t}}^{(\t{q,J})}:=\openone^{(\t{J})}\oplus\mathsf{R}_{\t{\omega_\t{P}t}}$.

\section{Tracking unknown signals with polynomial models}
\label{app:kin_models}
In the final section of the appendix, we discuss how to construct KF-based estimators for waveforms of unknown average dynamics 
after approximating their behaviour by means of the so-called \emph{polynomial models} and adequately augmenting the state space 
with waveform derivatives \cite{bar2004,vanTrees2013}. As discussed in the main text (see \eqnref{eq:dyns_pump_unknown}), we 
are interested in estimating waveforms, $\vB{q}_t$, which follow unknown dynamics, $\bar{\vB{q}}_t$, and experience fluctuations 
of known statistical properties, $\d \vB{w}_t^{(0)} := [\d w_q(t),\d w_p(t)]^T \sim\cN(0,\mB{Q}\,\d t)$, so that in the It\={o} form:
\BE
\d \vB{q}_t = \d \bar{\vB{q}}_t + \d \vB{w}_t^{(0)} = \dot{\bar{\vB{q}}}_t \d t + \d \vB{w}_t^{(0)}. \label{eq:dyns_pump_unknown_appendix} 
\EE
Abusing the It\={o} notation and including higher orders of $\d t$, we explicitly Taylor-expand the differential $\d \bar{\vB{q}}_t$ as follows
\BE
	\d \bar{\vB{q}}_t \approx \bar{\vB{q}}(t+\d t)-\bar{\vB{q}}(t) = \sum_{n=1}^{\infty}\frac{\bar{\vB{q}}_t^{(n)}} {n!} \d t^{n}, \label{eq:pm_expan}
\EE
where $\bar{\vB{q}}_t^{(n)}:=\frac{\d^n \bar{\vB{q}}_t}{\d t^n}$ denotes the $n$th time-derivative of the signal at time $t$.
By adopting the polynomial model one approximates the expansion \eref{eq:pm_expan} up to some order,  $l$, and assumes the noise term, 
$\d \vB{w}_t^{(0)}$, to originate solely from fluctuations of $\bar{\vB{q}}_t^{(l)}$. In particular, the evolution of the state vector $\vB{q}_t$ is 
then modelled as the solution of the following set of $l+1$ coupled (stochastic) differential equations:
\BE
\begin{array}{crcl}
n = 0,\; & ~\d \bar{\vB{q}}_t~ & = & {\bar{\vB{q}}}_t^{(1)} \d t \\
n = 1,\; & \d \bar{\vB{q}}_t^{(1)}   &= &{\bar{\vB{q}}}_t^{(2)} \d t \\
n = 2,\; & \d \bar{\vB{q}}_t^{(2)}   &= &{\bar{\vB{q}}}_t^{(3)} \d t \\
\vdots &   & \vdots & \\
n = l-1,\;  &\d \bar{\vB{q}}_t^{(l-1)}     &=  & {\bar{\vB{q}}}_t^{(l)} \d t\\
n = l,\;  &\d \bar{\vB{q}}_t^{(l)}     &=  & \d \vB{w}_t^{(l)}, \\
\end{array}
\label{ex:taylors}
\EE
where the stochastic fluctuations of $\bar{\vB{q}}_t^{(l)}$ are determined by an 
effective noise-term such that $\d \vB{w}_t^{(l)}\delta t^l = \d \vB{w}_t^{(0)}$.
The effective time-interval $\delta t$ relating the noise-strengths between the $n=0$ and
$n=l$ levels should be set by an educated guess \cite{bar2004}. However, in case of time-discrete measurement models---%
in particular our atomic sensor implementation---it is determined by the sampling period $\Delta$.
Crucially, within the polynomial model one treats all the derivatives in \eqnref{ex:taylors} as 
independent elements of the state vector. As a result, the state space of the quadrature must be enlarged, 
so that the new \emph{augmented quadrature-vector} reads
\BE
\vB{q}_t^\t{A}:=\vB{q}_t\oplus\vB{q}_t^{+}
\quad\text{with}\quad 
\vB{q}_t^{+}:=[\bar{\vB{q}}_t^{(1)},\bar{\vB{q}}_t^{(2)},\dots,\bar{\vB{q}}_t^{(l)}]^T.
\EE

In case of the atomic sensor implementation, we consider a polynomial model approximating the waveform up to $l=2$, i.e., the \emph{Wiener process accelerations model}
\cite{bar2004}, whose enlarged state space contains then also $\dot{{\vB{q}}}_t\equiv{{\vB{q}}}_t^{(\t{1})}$ and $\ddot{{\vB{q}}}_t\equiv{{\vB{q}}}_t^{(\t{2})}$ 
(in what follows, we drop the $\bar{\bullet}$-notation for simplicity). In such a case, we can write the dynamics of the augmented quadrature-vector as
\BE
\d \vB{q}_t^\t{A} = \mB{F}^{(\t{q,q^+})} \vB{q}_t^\t{A} \d t+ \d \vB{w}_t^{(\t A)}
\label{eq:dyn_quad_aug}
\EE
where for the ordering such that $\vB{q}_t^\t{A}=[q_t,p_t,\dot{q}_t,\dot{p}_t,\ddot{q}_t,\ddot{p}_t]^T$ the process $\mB{F}$-matrix reads
\BE
\mB{F}^{(\t{q,q^+})}  =
\left[\begin{array}{cccccc}
0 & 0 & 1 & 0 & 0 & 0\\
0 & 0 & 0 & 1 & 0 & 0\\
0 & 0 & 0 & 0 & 1 & 0\\
0 & 0 & 0 & 0 & 0 & 1\\
0 & 0 & 0 & 0 & 0 & 0\\
0 & 0 & 0 & 0 & 0 & 0
\end{array}\right]
\EE
coupling $\vB{q}_t$, $\dot{\vB{q}}_t$ and  $\ddot{\vB{q}}_t$ as prescribed by \eqnref{ex:taylors}. Moreover, the noise term in
\eqnref{eq:dyn_quad_aug} accordingly affects then only the second derivatives (accelerations), i.e., 
$\d \vB{w}_t^{(\t A)}=[0,0,0,0,\d \vB{w}_t^{(\ddot{\mathrm{q}})},\d \vB{w}_t^{(\ddot{\mathrm{p}})}]^T$ (see also \eqnref{eq:pm} of the main text).

Consequently, the dynamics of the \emph{augmented} state vector that contains now both the spin (J) and
augmented quadratures (A) degrees of freedom, i.e.,
\BE
\vB{x}_{t}
=
\vB{j}_{t} \oplus \vB{q}_{t}^{\t A}
=
\vB{j}_{t} \oplus \vB{q}_{t} \oplus \vB{q}_{t}^{+},
\label{eq:state_vector_track}
\EE
is described by the dynamical process \eref{eq:cd_model_state} (with $\mB{\Gamma}_{t}=0$, $\mB{G}_{t}=\openone$, as in the case of \appref{sec:KF_atoms}):
\BE
\d \vB{x}_t = \mB{F}_t \vB{x}_t \d t+ \d \vB{w}_t,
\EE 
where the augmented noise-term $\d \vB{w}_{t}$ and the process matrix $\mB{F}_t$ now, respectively, read:
\BE
\d \vB{w}_{t} =\d \vB{w}_t^{(\t J)} \oplus \d \vB{w}_t^{(\t A)}
, \qquad 
\mB{F}_t = \mB{F}_t ^{(\t{J,q})}\oplus\openone^{(\t{q^+})}\;+\;\openone^{(\t{J})}\oplus\mB{F}_t ^{(\t{q,q^+})},
\EE
and
\BE
\mB{F}_t ^{(\t{J,q})}=\left[\begin{array}{cc|cc}
-\frac{1}{T_{2}} & \omega_{\mathrm{L}} & 0 & 0\\
-\omega_{\mathrm{L}} & -\frac{1}{T_{2}} & \gP\cos\!\left(\omega_{\t{P}}t\right) & \gP\sin\!\left(\omega_{\t{P}}t\right)\\
\hline 0 & 0 & 0 & 0\\
0 & 0 & 0 & 0
\end{array}\right],
\EE
is the $\mB{F}$-matrix of \eqnref{eq:F_matrix} with $\kappa_q=\kappa_p=0$ that is determined for the atomic sensor by the coupled evolution 
of the input waveform and the ensemble spin, as specified by \eqnsref{eq:dyns_atoms}{eq:dyns_pump} of the main text (with $\kappa_q=\kappa_p=0$). 

On the other hand, the measurement process (i.e., the light-detection of atoms described by \eqnref{eq:I_meas} of the main text) and, hence, the observation 
model \eref{eq:obs_model_exp} remain the same as in the case of tracking fluctuating signals of known average form.
Finally, with both the dynamical and observation models at hand, the KF---in particular, the HKF introduced in \appref{sec:cd_model}---can be implemented in 
exactly analogous manner to \appref{sec:KF_atoms}, in order to now track noisy waveforms whose average dynamics is not known.

\end{document}